\documentclass[journal]{IEEEtran}
\usepackage{cite}
\usepackage{amsmath,amssymb,amsfonts}
\usepackage{algorithmic}
\usepackage{algorithm}
\usepackage{graphicx}
\usepackage{textcomp}
\usepackage{xcolor}
\usepackage{flushend}
\ifCLASSOPTIONcompsoc
\usepackage[caption=false,font=normalsize,labelfont=sf,textfont=sf]{subfig}
\else
 \usepackage[caption=false,font=footnotesize]{subfig}
\def\BibTeX{{\rm B\kern-.05em{\sc i\kern-.025em b}\kern-.08em
    T\kern-.1667em\lower.7ex\hbox{E}\kern-.125emX}}
\newcommand{\Deltaeq}{\ensuremath{\stackrel{\Delta}{=}}}

\title{Near-Field Localization with Antenna Arrays in the Presence of Direction-Dependent Mutual Coupling}

\author{Zohreh Ebadi,~\IEEEmembership{Student Member,~IEEE}, Amir Masoud Molaei,~\IEEEmembership{Senior~Member,~IEEE}, \\George C. Alexandropoulos,~\IEEEmembership{Senior~Member,~IEEE}, Muhammad Ali Babar Abbasi,~\IEEEmembership{Member,~IEEE}, \\Simon Cotton,~\IEEEmembership{Senior~Member,~IEEE}, Anvar Tukmanov, and Okan Yurduseven,~\IEEEmembership{Senior~Member,~IEEE}
\thanks{This work has been supported by Leverhulme Trust under Research Leadership Award under Grant RL-2019-019, British Telecom under UKRI Future Leaders Fellowship grant MR/T019980/1,  the SNS JU project TERRAMETA under the European Union’s Horizon Europe research and innovation program under Grant Agreement No 101097101, including top-up funding by UKRI under the UK government's Horizon Europe funding guarantee, and by the Department for the Economy of Northern Ireland under US Ireland R$\&$D Partnership grant no. USI 199.}
\thanks{Z. Ebadi, A. M. Molaei, M. A. B. Abbasi, S. Cotton, and O. Yurduseven are with the Centre for Wireless Innovation, School of Electronics, Electrical Engineering, and Computer Science, Queen's University Belfast, BT3~9DT Belfast, UK (e-mails: \{zebadi01, a.molaei, m.abbasi, simon.cotton, okan.yurduseven\}@qub.ac.uk).}
\thanks{G. C. Alexandropoulos is with the Department of Informatics and Telecommunications, National and Kapodistrian University of Athens, 15784 Athens, Greece and with the Department of Electrical and Computer Engineering Department, University of Illinois Chicago, Chicago, IL 60601, USA (e-mail: alexandg@di.uoa.gr).}
\thanks{A. Tukmanov is with the BT Labs, Adastral Park, BT Group, Ipswich, UK (e-mail: anvar.tukmanov@bt.com).}
}

\begin{document}
\maketitle
\begin{abstract}
Localizing near-field sources considering practical arrays is a recent challenging topic for next generation wireless communication systems. 
Practical antenna array apertures with closely spaced elements exhibit direction-dependent mutual coupling (MC), which can significantly degrade the performance localization techniques. A conventional method for near-field localization in the presence of MC is the three-dimensional (3D) multiple signal classification technique, which, however, suffers from extremely high computational complexity. Recently, two-dimensional (2D) search alternatives have been presented, exhibiting increased complexity still for direction-dependent MC scenarios. In this paper, we devise a low complexity one-dimensional (1D) iterative method based on an oblique projection operator (IMOP) that estimates direction-dependent MC and the locations of multiple near-field sources. The proposed method first estimates the initial direction of arrival (DOA) and MC using the approximate wavefront model, and then, estimates the initial range of one near-field source using the exact wavefront model. Afterwards, at each iteration, the oblique projection operator is used to isolate components associated with one source from those of other sources. The DOA and range of this one source are estimated using the exact wavefront model and 1D searches. Finally, the direction-dependent MC is estimated for each pair of the estimated DOA and range. The performance of the proposed near-field localization approach is comprehensively investigated and verified using both a full-wave electromagnetic solver and synthetic simulations. It is showcased that our IMOP scheme performs almost similarly to a state-of-the-art approach but with a $42$ times less computational complexity.
\end{abstract}


\section{Introduction}
The localization of devices and objects, network-connected or not, is important in many practical applications of array processing, such as wireless communication systems and vehicle localization~\cite {Tao2022,Molaei2022}.
Generally, localization algorithms can be classified based on the source location with respect to the localizer. To this end, there exist both far- and near-field localization algorithms~\cite{Luan2022,Zhang2023}. If the source is in the near-field region, known as the Fresnel region, the received wavefront from the near-field source is spherical. 
In this case, the received wavefront will be characterized using both direction of arrival (DOA) and range. In contrast, the wavefront is received as a plane wave when the source is located in the far-field region. Therefore, the far-field wavefront will be characterized by only the DOA~\cite{Position_aided, Jang2024,AVW22a,R_RIS,Ghazalian2024,Ghazalian2024b}. Hence, it is not feasible to localize near-field sources using algorithms developed for far-field source localization.

Over the recent decades, a significant number of algorithms have been developed for near-field source localization, focusing on DOA and range estimation, e.g., \cite{Guanghui2020,AbuShaban2021,Li2021,Cheng2022,RIS_near_field,gavras2023_hmimo_ISAC,gavras2024dma_1bit,Chen2023,Cheng2023,Xue2023,Gavras_SPAWC_2024}. These methods rely on system modeling, including the behavior of electromagnetic waves and the effects due to electromagnetic phenomena, such as the mutual coupling (MC) that refers to the electromagnetic interaction between the antenna elements of an array. The wavefront model, which considers the fact that the distance between the source and each element in the receiver array is not the same, is called the \textit{exact model}\cite{Ebadi2024,Chen2023,Friedlander2019}. Many near-field localization algorithms rely on a simplified exact model, called the \textit{approximated model}~\cite{AbuShaban2021,gavras2023_hmimo_ISAC,Ebadi2024a}, for ease of mathematical handling and computational feasibility. However, using the approximated model may lead to localization errors \cite{Ebadi2024,Friedlander2019}. For example, the existence of MC can reduce the accuracy of localization algorithms~\cite{Mohsen2023, Lan2023, Liu2023}. Based on the type of elements in the antenna array, the MC can be categorized as either direction-independent or direction-dependent \cite{Ge2020, Qi2019, Elbir2017, Zheng2021}. The former appears in arrays consisting of omnidirectional antennas. However, for practical arrays with non-omnidirectional antennas, the MC is usually direction-dependent~\cite{Elbir2017, Qi2019}. Near-field localization, while considering MC, has been treated in~\cite{Famoriji2021, Abedin2012, Xie2016, Wen2023, Chen2019}. The methods in~\cite{Xie2016, Wen2023, Chen2019} focused on mixed far- and near-field sources. Firstly, the DOA and MC of the far-field sources were estimated. Then, those estimations were used to compensate for the received signal and estimate the location of the near-field source, specifically, its DOA and range. However, those methods failed in scenarios where only the near-field sources exist in the presence of MC. In \cite{Famoriji2022,Famoriji2021, Abedin2012}, techniques aiming to estimate the location of near-field sources in the presence of direction-independent MC using, however, specific antenna arrays, namely, spherical and circular ring arrays, were developed. 

The rapid increase in the number of antennas in receiver arrays as well as the consideration of high-frequency bands expands the boundaries of the near-field region~\cite{Lu2023,NF_beam_tracking}. Hence, it is of paramount importance to devise localization schemes that can estimate all near-field sources in the presence of direction-dependent MC between the receive antenna elements, without relying on far-field sources to estimate and compensate for the MC. In our previous work~\cite{Ebadi2024a}, using the exact wavefront model, we presented two near-field source localization methods taking into account the direction-dependent MC among the receiver antenna array elements. The first method was a two-dimensional (2D) search method for near-field source localization (TSMNSL), with significantly less computational complexity than the three-dimensional (3D) multiple signal classification (MUSIC), which performs a 3D search for estimating DOA, range, and MC. The second method aimed to reduce TSMNSL's complexity and relied on an iterative technique based on a one-dimensional (1D) search.
However, that method could only estimate the location of a \textit{single} source. In this paper, we extend our second method in~\cite{Ebadi2024a} to estimate the location of several near-field sources using the oblique projection operator~\cite{Boyer2008}. Interestingly, the proposed method does not need any further process to pair the estimated DOA and range. The proposed method not only estimates the DOA and range of the near-field sources but also estimates the corresponding direction-dependent MCs. This method uses estimated MCs to improve the accuracy of location estimation in each iteration. The estimated MC can also be used for channel reconstruction problems. The main contributions of this paper are summarized as follows. 
\begin{itemize}
    \item We present a novel 1D iterative method based on oblique projection (IMOP) to estimate the direction-dependent MC and the location of multiple near-field sources. The proposed method has very low computational complexity employing only 1D searches, and utilizes the oblique projection operator to isolate components associated with one source from those of other sources in each iteration. 
    \item The presented analysis relies on the exact wavefront model, avoiding localization errors caused by the frequently used wavefront approximations.
    \item We derive the Cram$\Acute{\text{e}}$r-Rao lower bound (CRLB) for the DOA, range, and MC estimations using direction-dependent MC and the exact wavefront model.
    \item The performance of the presented IMOP method for near-field localization under direction-dependent MC is validated using both full-wave electromagnetic simulations, via the CST Microwave Studio~\cite{CSTStudio}, and synthetic simulations, via MATLAB.  
\end{itemize} 

\textit{Notation:} $\mathbf{I}_m$, $\mathbf{1}_m$, and $\mathbf{0}_{1\times m}$ represent the $m\times m$ identity matrix, the $m\times m$ all-ones matrix, and the ${1\times m}$ null vector. Additionally, $E\{\cdot\}$, $\text{Tr}\{\cdot\}$, $(\cdot)^T$, $(\cdot)^H$, and $\oplus$ stand for the statistical expectation, trace operator, transposition, Hermitian transposition, and the direct sum operator, respectively. Furthermore, $\frac{\partial \mathcal{Y}}{\partial y}$, $\hat{x}$, $\mathcal{R}(\cdot)$, and $[\cdot]_{ij}$ denote the first
derivative of $\mathcal{Y}$ with respect to $y$, the estimate of $x$, the range space of the bracketed matrix, and the $(i,j)$-th element of the matrix inside of the bracket, respectively. The letter $\jmath$ denotes the imaginary unit. $\text{Re}\{x\}$ returns the real part of the complex number $x$.

\section{System Model}\label{System_model_Section}
\begin{figure}[!t]
\centerline{\includegraphics[width=3.3in]{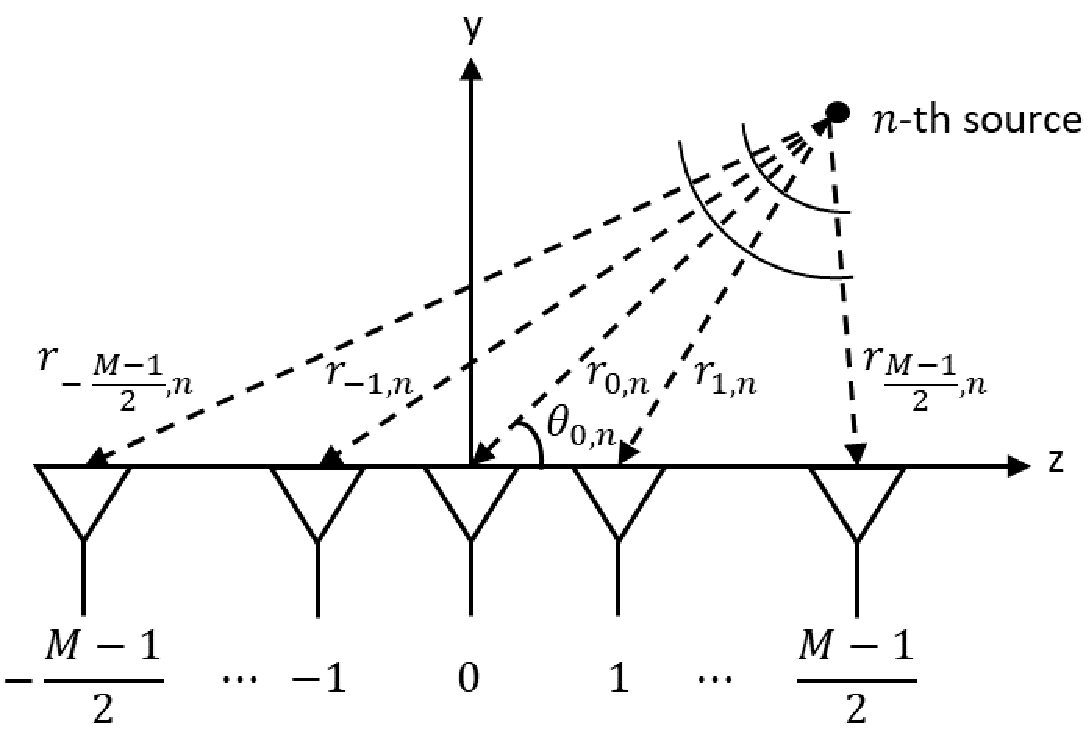}}
\caption{A uniform linear array geometry and the spherical wave propagation (including ranges and directions of arrival) from the $n$-th source located in the array's near-field region.}
\label{sourcelocation}
\end{figure}

Suppose that $N$ sources are located in the near-field of a uniform linear array (ULA) consisting of $M$ elements (suppose $M$ is an odd number), as shown in Fig.~\ref{sourcelocation}. 
The inter-element spacing in the ULA is $d={\lambda}/{2}$, where $\lambda$ is the signal wavelength. The signal from each $n$-th source ($n=1,2,\ldots,N$) impinging on the ULA with the DOA $\theta_{0,n}$ and range $r_{0,n}$ is considered to be a narrowband uncorrelated signal. In particular, $\theta_{0,n}$ is defined as the DOA of the signal $s_n(l)$ ($l=1,2,\ldots,L$ denotes the snapshot number) at the zeroth reference element in the ULA, and $r_{0,n}\in[0.62(D^3/\lambda)^\frac{1}{2},2D^2/\lambda]$~\cite{Molaei2022} is the distance between the $n$-th source and the reference element with $D\triangleq(M-1)d$ denoting the array aperture size. The signal received at each $l$-th snapshot can be mathematically expressed as follows~\cite{Ge2020}:
\begin{align}\label{system_model}
   \mathbf{y}(l) \triangleq \sum^N_{n=1}s_n(l)\mathbf{\Tilde{a}}(\theta_{0,n},r_{0,n},\mathbf{c}_n)+\mathbf{w}(l),
\end{align}
where $\mathbf{w}(l)\in \mathbb{C}^{M\times1}$ represents the additive noise vector at the $l$-th snapshot and $N$ is the total number of sources. In addition, the vector $\mathbf{\Tilde{a}}(\theta_{0,n},r_{0,n},\mathbf{c}_n)$ can be written as~\cite{Liao2012}:
\begin{align}\label{a_Tilda}
    \mathbf{\Tilde{a}}(\theta_{0,n},r_{0,n},\mathbf{c}_n)\triangleq\mathbf{C}(\theta_{0,n})\mathbf{a}(\theta_{0,n},r_{0,n}),
\end{align}
where $\mathbf{C}(\theta_{0,n})\in \mathbb{C}^{M\times M}$ is the direction-dependent MC matrix of the ULA at the $n$-th angle, given by~\cite{Ge2020,Elbir2017}:
\begin{align}
    \mathbf{C}(\theta_{0,n})=\text{Toeplitz}\{\mathbf{z}_n\},
\end{align}
where $\mathbf{z}_n\triangleq[\mathbf{c}_n^T\,\mathbf{0}_{1\times(M-Q)}]^T$ with $Q<M$ and $\mathbf{c}_n\triangleq[1\,c_{2,n}\,\cdots\,c_{(Q-1),n}]^T$ is the $Q\times1$ complex-valued vector of MC coefﬁcients due to the presence of the $n$-th near-field source.
The latter implies that, as the index distance between $\mathbf{c}_n$'s elements increases, the magnitude of the MC coefficients decreases. Beyond $Q$ inter-element spacings, the coupling coefficients become negligible~\cite{Ge2020}. Finally,
$\mathbf{a}(\theta_{0,n},r_{0,n})$ in~\eqref{a_Tilda} represents the $M\times 1$ steering vector with elements given for the exact near-field propagation model as follows~\cite{Friedlander2019}:
\begin{align}\label{steering_vector1}
    a_m(\theta_{0,n},r_{0,n}) = \frac{r_{0,n}}{r_{m,n}}e^{-\jmath\tau_{m,n}},
\end{align}
where $m=-\frac{M-1}{2},\ldots,-1,0,1,\ldots,\frac{M-1}{2}$. In this expression, $\tau_{m,n}$ denotes the phase difference between each $n$-th source and each $m$-th element in the ULA, which is given by:
\begin{align}\label{Tau}
    \tau_{m,n}= \frac{2\pi}{\lambda}(r_{m,n}-r_{0,n}),
\end{align}
with $r_{m,n}$ being the distance between the $n$-th source and the array's $m$-th element, which is easily obtained as follows~\cite{Friedlander2019}:
\begin{align}\label{distance}
    r_{m,n} = \sqrt{r^2_{0,n}+m^2d^2-2mdr_{0,n}\cos{\theta_{0,n}}}~.
\end{align}

In most near-field localization approaches (e.g.,~\cite{Zhang2018, Molaei2022, Guanghui2020}), an approximate version of $\mathbf{a}(\theta_{0,n},r_{0,n})$, called the approximation model, is used instead of~\eqref{steering_vector1}. To derive this approximation model, first, the far-field assumption is used, leading to the approximation $\frac{r_{0,n}}{r_{m,n}}\approx 1$ that results in the removal of the magnitude factor in \eqref{steering_vector1}. Then, using the second-order Taylor approximation, $r_{m,n}$ is written as follows:
\begin{align}\label{TaylorApproximation}
    r_{m,n}\approx &r_{0,k}\left(1+\frac{m^2d^2}{2r_{0,k}^2}-\frac{md}{r_{0,k}}\cos{\theta_{0,n}}-\frac{m^4d^4}{8r_{0,k}^4}\right.\nonumber\\
    &\left.-\frac{m^3d^3}{2r_{0,k}^3}\cos{\theta_{0,n}}-\frac{m^2d^2}{2r_{0,k}^2}\cos^2{\theta_{0,n}}\right).
\end{align}
Assuming that $d^3\ll r_{0,k}^3$ and $d^4\ll r_{0,k}^4$, \eqref{TaylorApproximation} can be further simplified as follows:
\begin{align}\label{TaylorApproximation2}
    r_{m,n}&\approx r_{0,k}\left(1-\frac{md}{r_{0,k}}\cos{\theta_{0,n}}+\frac{m^2d^2}{2r_{0,k}^2}\sin^2{\theta_{0,n}}\right).
\end{align}
Putting all above together, the approximate version for the exact model $a_m(\theta_{0,n}, r_{0,n})$ in~\eqref{steering_vector1} can be expressed as follows:
\begin{align}\label{steering_vector_nearfield}
    \bar{a}_m(\theta_{0,n},r_{0,n}) = e^{\jmath(\gamma_nm+\eta_n m^2)},
\end{align}
where we have used the definitions:
\begin{align}\begin{split}
    \gamma_n&\triangleq-\frac{2\pi d}{\lambda}\cos{\theta_{0,n}}, \\
    \eta_n&\triangleq\frac{\pi d^2}{\lambda r_{0,n}}\sin^2{\theta_{0,n}}.
    \end{split}
\end{align}

In this paper, we focus on the estimation of $\theta_{0,n}$ and $r_{0,n}$ $\forall$$n=1,2,\ldots,N$ in the presence of unknown direction-dependent MC using the measurements in~\eqref{system_model}. For this goal, we make the following assumptions:
 \begin{enumerate}
    \renewcommand{\labelenumi}{A\arabic{enumi})}
    \item The signals $\{s_n(l)\}_{n=1}^N$ are uncorrelated complex white Gaussian random processes with zero mean.
    \item The total number of sources $N$ in the vicinity of the $M$-element ULA receiver is known and holds $N<M$.
    \item The additive noise $\mathbf{w}(l)$ is a complex Gaussian random process with a zero mean and variance of $\sigma^2$ and is statistically independent of the sources’ signals.
    \item All DOAs are distinct, i.e., $\theta_{0,i} \neq \theta_{0,{i^{'}}}$ for $i\neq i^{'}$, and all ranges are also distinct, i.e., $r_{0,i} \neq r_{0,{i^{'}}}$ for $i\neq i^{'}$. Therefore, the rank of the array steering matrix $\mathbf{A}$ is $N$.
\end{enumerate}

\section{The TSMNSL Algorithm}\label{proposed_algorithms}
In this section, we present our TSMNSL algorithm for the estimation of the direction-dependent MC, as well as the DOAs and ranges of all near-field sources.

\subsection{Definition of the Transformation Matrix (TM)}
Starting from \eqref{a_Tilda}, $\mathbf{\Tilde{a}}(\theta_{0,n},r_{0,n})$ can be re-written as follows:
\begin{align}\label{Transform_defenition}
&\mathbf{\Tilde{a}}(\theta_{0,n},r_{0,n},\mathbf{c}_n)=\sum_{m=1}^{Q}\mathbf{E}_m\mathbf{a}(\theta_{0,n},r_{0,n})c_{m,n} \nonumber \\
    &=[\mathbf{E}_1\mathbf{a}(\theta_{0,n},r_{0,n}))\, \cdots \, \mathbf{E}_{Q}\mathbf{a}(\theta_{0,n},r_{0,n})][1\ c_{2,n}\ \cdots \ c_{Q,n}]^T \nonumber \\
    &=\mathbf{X}(\theta_{0,n},r_{0,n})\mathbf{c}_n,
\end{align}
where $\mathbf{X}(\cdot,\cdot)$ is the ${M\times Q}$ TM depending on both $\theta_{0,n}$ and $r_{0,n}$ and the matrix $\mathbf{E}_m$ is defined as:
\begin{align}
    [\mathbf{E}_m]_{ij}=
    \begin{cases}
      1, & [\mathbf{C}]_{ij}(\theta_{0,n})=c_{m,n}\\
      0, & \text{otherwise}
    \end{cases}.
\end{align}

\subsection{Estimation of the Sources' Parameters}
Following the system assumptions of Section~\ref{System_model_Section}, the covariance matrix of the received signal in \eqref{system_model} is computed as:
\begin{align}\label{covariance_exactequation}
    \mathbf{R}=E\{\mathbf{y}(l)\mathbf{y}^H(l)\}=\mathbf{\Tilde{A}}\mathbf{R}_s\mathbf{\Tilde{A}}^H+\sigma^2\mathbf{I}_M,
\end{align}
where $\mathbf{\Tilde{A}}\in \mathbb{C}^{M\times N}$ represents the array steering matrix having $\mathbf{\Tilde{a}}(\theta_{0,n},r_{0,n},\mathbf{c}_n)$ at its $n$-th column and $\mathbf{R}_s=E\{\mathbf{s}(l)\mathbf{s}^H(l)\}$ is the signal covariance matrix with vector $\mathbf{s}(l)$ consisting of the $N$ source signals at each $l$-th snapshot. In practice, for the $L$ measurements $\{\mathbf{y}(l)\}^L_{l=1}$ being available at the receiver, the covariance matrix of the received signal can be estimated as:
\begin{align}\label{covariance}
    \hat{\mathbf{R}}=\frac{1}{L}\sum_{l=1}^{L}\mathbf{y}(l)\mathbf{y}^H(l).
\end{align}
which can be eigendecomposed as $\hat{\mathbf{R}}=\mathbf{U}\mathbf{\Sigma}\mathbf{U}^H$,
where $\boldsymbol{\Sigma}$ includes its eigenvalues and $\mathbf{U}\triangleq[\mathbf{U}_s\ \mathbf{U}_w]$ is its eigenvector matrix; $\mathbf{U}_s\in \mathbb{C}^{M\times N}$ and $\mathbf{U}_w\in \mathbb{C}^{M\times (M-N)}$ contain the eigenvectors of the signal and noise subspaces, respectively. Then, the unknown sources' parameters, including DOAs, ranges, as well as the receiver's direction-dependent MC, can be jointly obtained by searching the $N$ peaks of the pseudo-spectrum of the 3D-MUSIC algorithm, which is given as:
\begin{align}\label{MUSIC_Spectrum}
    P(\theta_{0,n},r_{0,n},\mathbf{c}_n)\triangleq\frac{1}{\mathbf{\Tilde{a}}^H(\theta_{0,n},r_{0,n},\mathbf{c}_n)\textbf{U}_w\textbf{U}_w^H\mathbf{\Tilde{a}}(\theta_{0,n},r_{0,n},\mathbf{c}_n)}.
\end{align}
Note that, to search over this spectrum, a search interval for each of the unknowns $\theta_{0,n}$, $r_{0,n}$, and $\mathbf{c}$ needs to be defined. The search intervals for $\theta_{0,n}$ and $r_{0,n}$ can be defined within the ranges of $\theta_{0,n}\in [0,\pi]$ and $r_{0,n}\in[0.62(D^3/\lambda)^\frac{1}{2},2D^2/\lambda]$~\cite{Molaei2022}, respectively. However, defining the search area for $\mathbf{c}$ is difficult. This is due to the fact that $\mathbf{c}$ contains complex values. It is also noted that even if we are able to define all parameters' search spaces, the 3D-MUSIC method suffers from a high computational cost. To address this challenge, \eqref{Transform_defenition} can be substituted into \eqref{MUSIC_Spectrum} to express the 3D-MUSIC spectrum as:
\begin{align}\label{MUSIC_Spectrum3}
    P(\theta_{0,n},r_{0,n},\mathbf{c}_n)=\frac{1}{\textbf{c}_n^H\boldsymbol{\Omega}(\theta_{0,n},r_{0,n})\textbf{c}_n},   
\end{align}
where $\boldsymbol{\Omega}(\theta_{0,n},r_{0,n})\triangleq\mathbf{X}^H(\theta_{0,n},r_{0,n})\textbf{U}_w\textbf{U}_w^H\mathbf{X}(\theta_{0,n},r_{0,n})$. Finding the $N$ peaks of this function is equivalent to solving the following optimization problem for given $\theta_{0,n}$ and $r_{0,n}$:
\begin{align}\label{Optimization}
    \min_{\mathbf{c}_n}\,\,\textbf{c}_n^H\boldsymbol{\Omega}(\theta_{0,n},r_{0,n})\textbf{c}_n\quad\text{s.t.}\quad  \mathbf{e}_1^H\mathbf{c}_n=1,    
\end{align}
where $\mathbf{e}_1$ is the first column of the unitary matrix. To solve this optimization problem, we formulate its Lagrangian function, with $\beta$ being the Lagrange multiplier, as follows~\cite{Bazzi2016}:
\begin{align}\label{lagrangian}
    \mathcal{L}(\mathbf{c}_n,\beta)\triangleq\mathbf{c}_n^H\boldsymbol{\Omega}(\theta_{0,n},r_{0,n})\mathbf{c}_n-\beta(\mathbf{e}_1^H\mathbf{c}_n-1).
\end{align}
By equating $\frac{\partial \mathcal{L}(\mathbf{c}_n,\beta)}{\partial \mathbf{c}_n}$ to zero and using the constraint $\mathbf{e}_1^H\mathbf{c}_n=1$, the estimated DOA and range for all sources are given by:
\begin{align}\label{Algorithm1Solution}
    \{\hat{\theta}_{0,n},\hat{r}_{0,n}\}_{n=1}^N=\arg\max_{\theta_{0,n},r_{0,n}}\textbf{e}^H_1\boldsymbol{\Omega}^{-1}(\theta_{0,n},r_{0,n})\textbf{e}_1.    
\end{align}
Finally, the MC in the presence of the $n$-th source can be computed as follows:
\begin{align}\label{EstimatedMCAlgorithm1}
    \hat{\mathbf{c}}_n=\frac{{\boldsymbol{\Omega}^{-1}(\hat{\theta}_{0,n},\hat{r}_{0,n})}\textbf{e}_1}{\textbf{e}_1^H{\boldsymbol{\Omega}^{-1}(\hat{\theta}_{0,n},\hat{r}_{0,n})}\textbf{e}_1}.
\end{align}

It can be seen from \eqref{Algorithm1Solution} that the presented TSMNSL method requires 2D searches to estimate the DOA and range of all $N$ near-field sources. In the next section, we present a near-field localization method with less computational complexity than the TSMNSL algorithm.

\section{Low Complexity Near-Field Localization}\label{SecondAlgorithm}
The proposed IMOP algorithm, relying on the exact wavefront model in~\eqref{steering_vector1}, for estimating the direction-dependent MC as well as the DOAs and ranges of the near-field sources consists of the following steps. First, initial DOAs and the MC are estimated utilizing the approximate wavefront model, and the initial range of the sources is estimated using the exact wavefront model. Next, the oblique projection is used to extract single-source information from the covariance matrix corresponding to the previous steps' DOA, range, and MC estimates. Then, an 1D search is deployed to estimate the DOA of each source separately, which is followed by another 1D search to estimate the range of each source. Finally, the MC is estimated for each pair of estimated DOA and range. The details of all the above steps are given below.

\subsubsection{Initial Estimation of DOAs and MC}
We commence by using the approximation model for the steering vector via~\eqref{steering_vector_nearfield} to compute initial estimates for the DOAs of the sources. To this end, the product $\mathbf{C}(\theta_{0,n})\bar{\mathbf{a}}(\theta_{0,n},r_{0,n})$ (with $\bar{\mathbf{a}}(\theta_{0,n},r_{0,n})\triangleq[\bar{a}_1(\theta_{0,n},r_{0,n})\,\cdots\,\bar{a}_M(\theta_{0,n},r_{0,n})]^T$ through the elements in~\eqref{steering_vector_nearfield}) can be written as follows:  
\begin{align} \label{Transformation_Approx}
    &\mathbf{C}(\theta_{0,n})\bar{\mathbf{a}}(\theta_{0,n},r_{0,n}) \nonumber \\
        &=\begin{bmatrix}
            g_{1} & c_{1} g_{2} & c_{2} g_{3} & \cdots & c_{M-1} g_{M}\\
             c_{1} g_{1} & g_{2} & c_{1} g_{3} & \cdots & c_{M-2} g_{M}\\
            \vdots & \vdots & \vdots & \vdots & \vdots\\
            c_{M-1} g_{1} & c_{M-2} g_{2} & c_{M-3} g_{3} & \cdots & g_{M}
        \end{bmatrix}\begin{bmatrix}
            v_1\\
            v_2\\
            \vdots\\
            v_M
        \end{bmatrix} \nonumber \\
        &=\mathbf{B}(\theta_{0,n},\mathbf{c}_n)\mathbf{v}(\theta_{0,n},r_{0,n}),
\end{align}
where ${v}_m(\theta_{0,n},r_{0,n})\triangleq e^{\jmath\eta_n m^2}$ and ${g}_m(\theta_{0,n})\triangleq e^{\jmath\gamma_nm}$ $\forall$$m=1,2,\ldots,M$.
For simplicity, we next define $\mathbf{B}_n\Deltaeq \mathbf{B}(\theta_{0,n},\mathbf{c}_n)$ and $\mathbf{v}_n\Deltaeq\mathbf{v}(\theta_{0,n},r_{0,n})$. By inserting \eqref{Transformation_Approx} into the  3D-MUSIC spectrum in \eqref{MUSIC_Spectrum}, yields the following estimation problem for the MC as well as the DOAs and ranges of all $N$ sources:
\begin{align}\label{Algorithm1_Solution2}
&\{\hat{\theta}_{0,n},\hat{r}_{0,n},\hat{\mathbf{c}}_n\}_{n=1}^N= \arg\min_{\theta_{0,n},r_{0,n},\mathbf{c}_n}\mathbf{v}_n^H\mathbf{B}_n^H\textbf{U}_w\textbf{U}^H_w\mathbf{B}_n\textbf{v}_n^H.    
\end{align}

To obtain an initial estimation for the MC and DOA of each $n$-th near-field source, i.e., $\hat{\mathbf{c}}_n^{(0)}$ and $\hat{\theta}_{0,n}^{(0)}$, we first focus on the following simplified form of \eqref{Algorithm1_Solution2}'s estimation problem:
\begin{align}\label{Algorithm1_Solution3}\begin{split}
\{\hat{\theta}_{0,n}^{(0)},\hat{\mathbf{c}}_n^{(0)}\}_{n=1}^N=\arg\min_{\theta_{0,n},\mathbf{c}_n}\mathbf{B}_n^H\textbf{U}_w\textbf{U}^H_w\mathbf{B}_n.    
    \end{split}
\end{align}
Following the same steps with the derivation of \eqref{Transform_defenition}, deduces to $\mathbf{B}_n=\bar{\mathbf{X}}(\theta_{0,n})\mathbf{c}_n$, where $\bar{\mathbf{X}}\triangleq[\mathbf{E}_1\mathbf{g}(\theta_{0,n}))\, \cdots \, \mathbf{E}_{Q}\mathbf{g}(\theta_{0,n})]$ (with ${\mathbf{g}}(\theta_{0,n})\triangleq[g_1(\theta_{0,n})\,\cdots\,g_M(\theta_{0,n})]^T$), which then yields the reformulation:
\begin{align}\label{Algorithm1_Solution4}
\{\hat{\theta}_{0,n}^{(0)},\hat{\mathbf{c}}_n^{(0)}\}_{n=1}^N=
\arg\min_{\theta_{0,n},\mathbf{c}_n}\textbf{c}_n^H{\bar{\mathbf{X}}}^{H}(\theta_{0,n})\textbf{U}_w\textbf{U}^H_w\bar{\mathbf{X}}(\theta_{0,n})\textbf{c}_n.  
\end{align}
To estimate each DOA, the following optimization problem needs to be solved for a given $\theta_{0,n}$:
\begin{align}\label{optimization4}
    \min_{\mathbf{c}_n} \mathbf{c}_n^H\bar{\boldsymbol{\Omega}}(\theta_{0,n})\mathbf{c}_n \quad
    \text{s.t.}\quad\mathbf{e}^H_1\mathbf{c}_n=1,
\end{align}
where $\bar{\boldsymbol{\Omega}}(\theta_{0,n})\triangleq\bar{\mathbf{X}}(\theta_{0,n})\textbf{U}_w\textbf{U}^H_w\bar{\mathbf{X}}(\theta_{0,n})$. Hence, $\hat{\theta}^{(0)}_{0,n}$ and initial estimate of MC correspond to $n$-th source, i.e., $\hat{\mathbf{c}}_n^{(0)}$, are computed as follows:
\begin{equation}\label{Initial_DOA}
    \{\hat{\theta}^{(0)}_{0,n}\}_{n=1}^N=\arg\max_{\theta_{0,n}}\mathbf{e}^H_1\bar{\boldsymbol{\Omega}}^{-1}(\theta_{0,n})\mathbf{e}_1,
\end{equation}
\begin{align}\label{Initial_MC}
    \hat{\mathbf{c}}_n^{(0)}=\frac{{\bar{\boldsymbol{\Omega}}}^{-1}(\hat{\theta}^{(0)}_{0,n})\textbf{e}_1}{\textbf{e}_1^H{\bar{\boldsymbol{\Omega}}}^{-1}(\hat{\theta}^{(0)}_{0,n})\textbf{e}_1}.
\end{align}
It is noted that, in these initial estimations for the MC and DOAs of all $N$ near-field sources, the approximate model of the received signal was used.

\subsubsection{Initial Estimation of Ranges} The initial estimation of the range, e.g., $\hat{r}_{0,n}^{(0)}$ for the $n$-th source corresponding to $\hat{\theta}^{(0)}_{0,n}$, can be found  from the solution of the following problem: 
\begin{equation}\label{Initial_r}
    \hat{r}_{0,n}^{(0)}=\arg\max_{r_{0,n}}\mathbf{e}^H_1\boldsymbol{\Omega}^{-1}(\hat{\theta}_{0,n},r_{0,n})\mathbf{e}_1,
\end{equation}
where $\boldsymbol{\Omega}(\hat{\theta}_{0,n},r_{0,n})$ is generated using the exact wavefront model for the received signal. This procedure is used for the initial estimation of all $N$ sources' ranges. 

\subsubsection{Iterative Section}\label{IterativeSection} 
Since the estimations $\{\hat{\theta}_{0,n}^{(0)}\}_{n=1}^N$ via \eqref{Initial_DOA} rely on the approximation model of the received wavefront, errors may occur which will also propagate to the estimations $\{\hat{\mathbf{c}}_n^{(0)}\}^N_{n=1}$ and $\{\hat{r}_{0,n}^{(0)}\}_{n=1}^N$ via \eqref{Initial_MC} and \eqref{Initial_r}, respectively. We now use the exact model of the received wavefront to improve these estimations in an iterative manner. In addition, we deploy oblique projection to isolate one signal from the other ones at each iteration. We also assume that, at each iteration, the DOA of one source is unknown while the DOAs of the other $N-1$ sources are known. Note that the error caused by using the approximate model in estimating the initial values can be more negligible than the errors caused by the random determination of initial values for the estimations.

Let $\mathbf{\Tilde{a}}_n\Deltaeq \mathbf{\Tilde{a}}(\theta_{0,n},r_{0,n},\mathbf{c}_n)$ be the $M\times 1$ array steering vector corresponding to the $n$-th source and $\mathbf{\Tilde{A}}_n$ denote the  $M\times (N-1)$ array steering matrix without the column $\mathbf{\Tilde{a}}_n$. The range space of $\mathbf{\Tilde{A}}$ can be expressed as follows:
\begin{align}\label{RangeSpace}
   \mathcal{R}(\mathbf{\Tilde{A}})=\mathcal{R}(\mathbf{\Tilde{a}}_n)\oplus\mathcal{R}(\mathbf{\Tilde{A}}_n), 
\end{align}
where $\mathcal{R}(\mathbf{\Tilde{a}}_n)$ and $\mathcal{R}(\mathbf{\Tilde{A}}_n)$ denote the range spaces of $\mathbf{\Tilde{a}}_n$ and $\mathbf{\Tilde{A}}_n$, respectively. In each $i$-th estimation iteration, $\mathbf{\Tilde{a}}_n^{(i)}$ is considered as the steering vector of the unknown parameters $(\theta_{0,n}^{(i)},r_{0,n}^{(i)},\hat{\mathbf{c}}_n^{(i)})$ and $\mathbf{\Tilde{A}}_n^{(i-1)}$ is the steering matrix of $N-1$ sources whose MC coefficients, DOAs, and ranges were estimated at the previous $(i-1)$-th iteration. Therefore, at each $i$-th iteration, the oblique projector $E_{{\mathbf{\Tilde{A}}_n^{(i-1)}}{\mathbf{\Tilde{a}}_n^{(i)}}}$ that projects onto the space $\mathcal{R}(\mathbf{\Tilde{A}}_n^{(i-1)})$ along a direction parallel to the space $\mathcal{R}(\mathbf{\Tilde{a}}_n^{(i)})$ is given as follows:
\begin{equation}\label{Oblique_Projection}
    E_{{\mathbf{\Tilde{A}}_n^{(i-1)}}{\mathbf{\Tilde{a}}_n^{(i)}}}={\mathbf{\Tilde{A}}_n^{(i-1)}}({\mathbf{\Tilde{A}}_n}^{H^{(i-1)}}\mathbf{P}_{\mathbf{\Tilde{a}}_n^{(i)}}^\perp \mathbf{\Tilde{A}}_n^{(i-1)})^{-1}{\mathbf{\Tilde{A}}_n}^{H^{(i-1)}}\mathbf{P}^\perp ,   
\end{equation}
where $\mathbf{P}^\perp_{\mathbf{\Tilde{a}}_n^{(i)}}$ represents the orthogonal projector onto the null space of the $n$-th source, which is obtained as:
\begin{align}
    \mathbf{P}^\perp_{\mathbf{\Tilde{a}}_n^{(i)}}=\mathbf{I}_M-\mathbf{\Tilde{a}}_n^{(i)}[\mathbf{\Tilde{a}}_n^{H^{(i)}}\mathbf{\Tilde{a}}_n^{(i)}]^{-1}\mathbf{\Tilde{a}}_n^{H^{(i)}}.
\end{align}
Note that the oblique projection has the following properties:
\begin{equation}\label{ObliqueProperties}
    E_{{\mathbf{\Tilde{A}}_n^{(i-1)}}{\mathbf{\Tilde{a}}_n^{(i)}}}\mathbf{\Tilde{A}}_n^{(i-1)}=\mathbf{\Tilde{A}}_n^{(i-1)}\ ,\ E_{{\mathbf{\Tilde{A}}_n^{(i-1)}}{\mathbf{\Tilde{a}}_n^{(i)}}}\mathbf{\Tilde{a}}_n^{(i)}=\mathbf{0}_{M\times 1}.
\end{equation}

Based on the latter definitions of $\mathbf{\Tilde{A}}_n$ and $\mathbf{\Tilde{a}}_n$, the received signal model in \eqref{system_model} can be re-written as follows:
\begin{align}\label{system_model_twoSection}
   \mathbf{y}(l) = \mathbf{\Tilde{a}}_n s_n(l)+\mathbf{\Tilde{A}}_n \mathcal{S}_n(l)+\mathbf{w}(l),
\end{align}
where $s_n(l)$ denotes the signal transmitted by the $n$-th source and $\mathcal{S}_n(l)\triangleq[s_1(l) \cdots s_{n-1}(l) \ s_{n+1}(l) \cdots s_N(l)]^T$ includes the signals transmitted by all other sources. Applying the oblique projection to \eqref{system_model_twoSection} at each $i$-th iteration, while using its properties summarized in~\eqref{ObliqueProperties}, deduces to:
\begin{align}\label{system_model_Oblique}
   \Bar{\mathbf{y}}^{(i)}(l) &= (\mathbf{I}_M-E_{{\mathbf{\Tilde{A}}_n^{(i-1)}}{\mathbf{\Tilde{a}}_n^{(i)}}})\mathbf{y}(l) \nonumber \\ 
   &=\mathbf{\Tilde{a}}_n s_n(l)+(\mathbf{I}_M-E_{{\mathbf{\Tilde{A}}_n^{(i-1)}}{\mathbf{\Tilde{a}}_n^{(i)}}})\mathbf{w}(l),
\end{align}
which implies that only information of one single source is included at the $i$-th iteration. Then, the covariance matrix of $\Bar{\mathbf{y}}^{(i)}(l)$ at this $i$-th iteration can be computed as follows:
\begin{align}\label{covariance_Oblique}
    \mathbf{\Bar{R}}^{(i)}=\frac{1}{L}\sum_{l=1}^{L}\mathbf{\Bar{y}}^{(i)}(l){\mathbf{\Bar{y}}}^{H^{(i)}}(l).
\end{align}
By eigendecomposing \eqref{covariance_Oblique}, the noise subspace $\tilde{\mathbf{U}}_w^{(i)}$ can be extracted, which is used to formulate the following matrix: 
\begin{align}\label{Omega_bar} 
\tilde{\boldsymbol{\Omega}}(\theta_{0,n}^{(i)},r_{0,n}^{(i)})=\mathbf{X}^H(\theta_{0,n}^{(i)},r_{0,n}^{(i)})\tilde{\textbf{U}}_w^{(i)}\tilde{\textbf{U}}_w^{{H^{(i)}}}\mathbf{X}(\theta_{0,n}^{(i)},r_{0,n}^{(i)}).
\end{align}
Using this matrix, the DOA of each $n$-th source at each $i$-th algorithmic iteration can be estimated as follows:
\begin{align}\label{Estimated_DOA}
    \hat{\theta}_{0,n}^{(i)}=\arg\max_{\theta_{0,n}}\mathbf{e}^H_1{\tilde{\boldsymbol{\Omega}}}^{-1}(\theta_{0,n},\hat{r}_{0,n}^{(i-1)})\mathbf{e}_1.
\end{align}
The corresponding range for this source is then estimated as:
\begin{align}\label{Estimated_range}
\hat{r}_{0,n}^{(i)}=\arg\max_{r_{0,n}}\mathbf{e}^H_1{\tilde{\boldsymbol{\Omega}}}^{-1}(\hat{\theta}_{0,n}^{(i)},r_{0,n})\mathbf{e}_1.
\end{align}
Finally, using all estimated DOAs and ranges, each $n$-th vector with MC coefficients can be estimated as follows:
\begin{align}\label{EstimatedMC_Algorithm2}
    \hat{\mathbf{c}}_n^{(i)}=\frac{{\tilde{\boldsymbol{\Omega}}^{-1}}(\hat{\theta}_{0,n}^{(i)},\hat{r}_{0,n}^{(i)})\textbf{e}_1}{\textbf{e}_1^H{\tilde{\boldsymbol{\Omega}}}^{-1}(\hat{\theta}_{0,n}^{(i)},\hat{r}_{0,n}^{(i)})\textbf{e}_1}.
\end{align}
If $|\hat{\theta}^{(i)}_{0,n}-\hat{\theta}^{(i-1)}_{0,n}|<\varepsilon$ with $\varepsilon$ being a small positive number, then $\hat{\theta}^{(i)}_{0,n}$, $\hat{r}^{(i)}_{0,n}$, and $\hat{\mathbf{c}}_n^{(i)}$ are considered the final estimates for DOA, range, and MC coefficients, respectively. The algorithmic iterations terminate if the latter condition is met; otherwise, a new iteration of estimations takes place. 
\begin{algorithm}[!t]
\caption{The Proposed IMOP Estimation Method}
\label{AlgorithmIMOP}
\resizebox{\columnwidth}{!}{%
\begin{minipage}{\columnwidth}
\begin{algorithmic}
    \STATE \textbf{Input:} $M$, $N$, $d$, and $\mathbf{y}(l)$.
    \STATE \textbf{Output:} $\{\hat{\theta}_{0,n},\ \hat{r}_{0,n},\ \hat{\mathbf{c}}_n\}_{n=1}^N$
    \STATE \hrulefill
    \STATE Compute $\{\hat{\theta}^{(0)}_{0,n}\}_{n=1}^N$, $\{\hat{\mathbf{c}}^{(0)}_n\}_{n=1}^N$, and $\{\hat{r}^{(0)}_{0,n}\}_{n=1}^N$ from \eqref{Initial_DOA}, \eqref{Initial_MC}, and \eqref{Initial_r}, respectively.
    \STATE Set iteration counter as $i=1$.
    \REPEAT
        \STATE Compute $\{\hat{\theta}^{(i)}_{0,n}\}_{n=1}^N$, $\{\hat{r}^{(i)}_{0,n}\}_{n=1}^N$, and $\{\hat{\mathbf{c}}_n^{(i)}\}^N_{n=1}$ from \eqref{Estimated_DOA}, \eqref{Estimated_range}, and \eqref{EstimatedMC_Algorithm2}, respectively.
        \STATE Set $i=i+1$.
    \UNTIL{$|\hat{\theta}^{(i)}_{0,n}-\hat{\theta}^{(i-1)}_{0,n}|<\varepsilon$}.
\end{algorithmic}
\end{minipage}}
\end{algorithm}

The proposed near-field multi-source localization method requires 1D searches to estimate all sources' DOAs and the ranges. In addition, the estimates $\hat{\theta}_{0,n}$ and $\hat{r}_{0,n}$ for each $n$-th source are automatically paired, hence, no further processing is required. Moreover, the proposed method does not consider initial values for the unknowns based on a guess (or randomly), which can increase the speed and accuracy of the estimation convergence. The iteration steps for the proposed IMOP method are given in Algorithm~\ref{AlgorithmIMOP}.

\subsection{Computational Complexities of TSMNSL and IMOP}\label{ComputationalComplexity}
The computational complexity of the eigendecomposition of the covariance matrix is $\mathcal{O}(M^3)$. Furthermore, the 2D searches for the DOA, and range estimations require the complexity of $\mathcal{O}(\mathcal{H}\mathcal{U}Q^2)$, where $\mathcal{H}=\frac{180}{\delta_\theta}$, $\mathcal{U}=\frac{\frac{2D^2}{\lambda}-\sqrt{0.62\frac{D^3}{\lambda}}}{\delta_r}$, and $\delta_\theta$ represents the step size for the DOA search, while $\delta_r$ represents the step size for the range search. Therefore, the overall computational complexity of the TSMNSL method is approximately $\mathcal{O}(M^3+\mathcal{H}\mathcal{U}Q^2)$.

For the IMOP method, the computational complexity of the initial estimations for the DOAs and ranges is approximately $\mathcal{O}(M^3+\mathcal{H}Q^2+\mathcal{U}Q^2)$. Furthermore, the 1D searches at each estimation iteration need approximately complexity of $\mathcal{O}(NM^3+N\mathcal{H}Q^2+N\mathcal{U}Q^2)$. Unlike the TSMNSL method, IMOP does not require a 2D search to estimate DOAs and ranges. Hence, its computational complexity is lower than that of the TSMNSL method.

\section{Estimation Analysis}
The CRLB provides a lower bound on the variance of any unbiased estimator and has been derived for near-field source localization in various relevant works~\cite{Weiss1993,Grosicki2005,Bayesian_CRB_ISAC}. In this section, we present the CRLB for the received system model in \eqref{system_model}. To this end, the CRLB is given by the inverse of the Fisher information matrix (FIM), i.e., $\text{CRLB}\Deltaeq\text{FIM}^{-1}$. Each $(n,n^{'})$-th element of the FIM for a parameter vector $\boldsymbol{\alpha}\Deltaeq[\boldsymbol{\theta}\ \mathbf{r}\ \mathbf{c}_1\ \cdots\ \mathbf{c}_N]$, with $\boldsymbol{\theta}\triangleq[\theta_{0,1}\,\theta_{0,2}\ \cdots\ \theta_{0,N}]$ and $\mathbf{r}\triangleq[r_{0,1}\,r_{0,2}\ \cdots\ r_{0,N}]$, is given by~\cite{kay}:
\begin{align}\label{FIM}
    [\text{FIM}]_{nn^{'}}=L\text{Tr}\left\{\frac{\partial \mathbf{R}}{\partial \boldsymbol{\alpha}_{n}}\mathbf{R}^{-1}\frac{\partial \mathbf{R}}{\partial \boldsymbol{\alpha}_{n^{'}}}\mathbf{R}^{-1}\right\}.
\end{align}

\subsection{FIM of the DOA Parameter}\label{FIMDOA}
The first derivative of the covariance matrix $\mathbf{R}$ given in \eqref{covariance_exactequation} with respect to $\theta_{0,n}$ is obtained as follows:
\begin{align}\label{DerivativeRrespectTheta}
    \frac{\partial\mathbf{R}}{\partial\theta_{0,n}}=\frac{\partial\Tilde{\mathbf{A}}}{\partial\theta_{0,n}}\mathbf{R}_s\Tilde{\mathbf{A}}^H+\Tilde{\mathbf{A}}\mathbf{R}_s\frac{\partial\Tilde{\mathbf{A}}^H}{\partial\theta_{0,n}},
\end{align}
By substituting this expression in \eqref{FIM}, the FIM of the DOA parameter can be expressed as:
\begin{align}\label{FIM_Theta}
    &[\text{FIM}(\boldsymbol{\theta})]_{nn^{'}}=L\text{Tr}\left\{\left(\frac{\partial\Tilde{\mathbf{A}}}{\partial\theta_{0,n}}\mathbf{R}_s\Tilde{\mathbf{A}}^H+\Tilde{\mathbf{A}}\mathbf{R}_s\frac{\partial\Tilde{\mathbf{A}}^H}{\partial\theta_{0,n}}\right)\mathbf{R}^{-1}\right.\nonumber\\
    &\left.\left(\frac{\partial\Tilde{\mathbf{A}}}{\partial\theta_{0,n^{'}}}\mathbf{R}_s\Tilde{\mathbf{A}}^H+\Tilde{\mathbf{A}}\mathbf{R}_s\frac{\partial\Tilde{\mathbf{A}}^H}{\partial\theta_{0,n^{'}}}\right)\mathbf{R}^{-1}\right\}\nonumber\\
    &=2L\text{Re}\left\{\text{Tr}\left\{\mathbf{R}^{-1}\frac{\partial\Tilde{\mathbf{A}}}{\partial\theta_{0,n}}\mathbf{R}_s\Tilde{\mathbf{A}}^H\mathbf{R}^{-1}\frac{\partial\Tilde{\mathbf{A}}}{\partial\theta_{0,n^{'}}}\mathbf{R}_s\Tilde{\mathbf{A}}^H\right\}\right.\nonumber\\
    &\left.+\text{Tr}\left\{\mathbf{R}^{-1}\frac{\partial\Tilde{\mathbf{A}}}{\partial\theta_{0,n}}\mathbf{R}_s\Tilde{\mathbf{A}}^H\mathbf{R}^{-1}\Tilde{\mathbf{A}}\mathbf{R}_s\frac{\partial\Tilde{\mathbf{A}}^H}{\partial\theta_{0,n^{'}}}\right\}\right\},
\end{align}
where $\frac{\partial\Tilde{\mathbf{A}}}{\partial\theta_{0,n}}=[\mathbf{0}_{M\times 1}\,\cdots\,\frac{\partial\Tilde{\mathbf{a}}(\theta_{0,n},r_{0,n})}{\partial\theta_{0,n}}\,\cdots\,\mathbf{0}_{M\times 1}]$ and $\frac{\partial\Tilde{
\mathbf{a}}(\theta_{0,n},r_{0,n})}{\partial\theta_{0,n}}=\frac{\partial \mathbf{C}(\theta_{0,n})}{\partial \theta_{0,n}}\mathbf{a}(\theta_{0,n},r_{0,n})+\mathbf{C}(\theta_{0,n})\frac{\partial
\mathbf{a}(\theta_{0,n},r_{0,n})}{\partial\theta_{0,n}}$. Furthermore, the derivative $\frac{\partial \mathbf{a}(\theta_{0,n},r_{0,n})}{\partial \theta_{0,n}}$ is computed as:
\begin{align}\label{d_theta}
   &\frac{\partial \mathbf{a}(\theta_{0,n},r_{0,n})}{\partial \theta_{0,n}}\nonumber\\
    &=-\left(\frac{1}{r_{m,n}}+\jmath\frac{2\pi}{\lambda}\right)md\frac{r_{0,n}}{r_{m,n}}\sin(\theta_{0,n})\mathbf{a}(\theta_{0,n},r_{0,n}).
\end{align}

\subsection{FIM of the Range Parameter}
Similarly, the first derivative of $\mathbf{R}$ in \eqref{covariance_exactequation} with respect to $r_{0,n}$ can be calculated as follows:
\begin{align}\label{DerivativeRrespectrange}
    \frac{\partial\mathbf{R}}{\partial r_{0,n}}=\frac{\partial\Tilde{\mathbf{A}}}{\partial r_{0,n}}\mathbf{R}_s\Tilde{\mathbf{A}}^H+\Tilde{\mathbf{A}}\mathbf{R}_s\frac{\partial\Tilde{\mathbf{A}}^H}{\partial r_{0,n}}.
\end{align}
where $\frac{\partial\Tilde{\mathbf{A}}}{\partial r_{0,n}}=[\mathbf{0}_{M\times 1}\,\cdots\,\frac{\partial\Tilde{\mathbf{a}}(\theta_{0,n},r_{0,n})}{\partial r_{0,n}}\,\cdots\,\mathbf{0}_{M\times 1}]$. By substituting this expression into \eqref{FIM}, the FIM of the range parameter is obtained as:
\begin{align}\label{FIM_range}
    &[\text{FIM}(\mathbf{r})]_{nn^{'}}=\nonumber\\
    &2L\text{Re}\left\{\text{Tr}\left\{\mathbf{R}^{-1}\frac{\partial\Tilde{\mathbf{A}}}{\partial r_{0,n}}\mathbf{R}_s\Tilde{\mathbf{A}}^H\mathbf{R}^{-1}\frac{\partial\Tilde{\mathbf{A}}}{\partial r_{0,n^{'}}}\mathbf{R}_s\Tilde{\mathbf{A}}^H\right\}\right.\nonumber\\
    &\left.+\text{Tr}\left\{\mathbf{R}^{-1}\frac{\partial\Tilde{\mathbf{A}}}{\partial r_{0,n}}\mathbf{R}_s\Tilde{\mathbf{A}}^H\mathbf{R}^{-1}\Tilde{\mathbf{A}}\mathbf{R}_s\frac{\partial\Tilde{\mathbf{A}}^H}{\partial r_{0,n^{'}}}\right\}\right\},
\end{align}
where we have used the following derivation:
\begin{align}\label{d_r}
    &\frac{\partial \mathbf{a}(\theta_{0,n},r_{0,n})}{\partial r_{0,n}}=\left(\frac{1}{r_{0,n}}+\jmath\frac{2\pi}{\lambda}\right)\mathbf{a}(\theta_{0,n},r_{0,n})\nonumber\\
    &-\left(\frac{1}{r_{m,n}}+\jmath\frac{2\pi}{\lambda}\right)\left(\frac{r_{0,n}-md\cos(\theta_{0,n})}{r_{m,n}}\right)\mathbf{a}(\theta_{0,n},r_{0,n}).
\end{align}

\subsection{FIM of the MC Parameter}
We consider, for simplicity, that the MC is given as $\mathbf{C}(\theta_{0,n})=\text{Toeplitz}\{[1\,\zeta_1\,\cdots\,\zeta_Q\,\mathbf{0}_{1\times (M-Q)}]^T\}$ where $\zeta_n\Deltaeq c_{m,n}$.  The first derivative of $\mathbf{R}$ in \eqref{covariance_exactequation} with respect to $\zeta_n$ is:
\begin{align}\label{DerivativeRrespectMC}
    \frac{\partial\mathbf{R}}{\partial \zeta_n}=\frac{\partial\Tilde{\mathbf{A}}}{\partial \zeta_n}\mathbf{R}_s\Tilde{\mathbf{A}}^H+\Tilde{\mathbf{A}}\mathbf{R}_s\frac{\partial\Tilde{\mathbf{A}}^H}{\partial \zeta_n}.
\end{align}
By using the derivations $\frac{\partial\Tilde{\mathbf{A}}}{\partial \zeta_n}=\frac{\partial \mathbf{C}(\theta_{0,n})}{\partial \zeta_n}\mathbf{a}(\theta_{0,n},r_{0,n})$ and $\frac{\partial \mathbf{C}(\theta_{0,n})}{\partial \zeta_n}=\text{Toeplitz}\{[\mathbf{0}_{1\times(n-1)}\ \mathbf{1}_1\ \mathbf{0}_{1\times(M-n)}]^T\}$, the FIM of the MC parameter is obtained as follows:
\begin{align}\label{FIM_MC}
    &[\text{FIM}(\mathbf{C}(\theta_{0,n}))]_{nn^{'}}\nonumber\\
    &=2L\text{Re}\left\{\text{Tr}\left\{\mathbf{R}^{-1}\frac{\partial\Tilde{\mathbf{A}}}{\partial \zeta_n}\mathbf{R}_s\Tilde{\mathbf{A}}^H\mathbf{R}^{-1}\frac{\partial\Tilde{\mathbf{A}}}{\partial \zeta_n^{'}}\mathbf{R}_s\Tilde{\mathbf{A}}^H\right\}\right.\nonumber\\
    &\left.+\text{Tr}\left\{\mathbf{R}^{-1}\frac{\partial\Tilde{\mathbf{A}}}{\partial \zeta_n}\mathbf{R}_s\Tilde{\mathbf{A}}^H\mathbf{R}^{-1}\Tilde{\mathbf{A}}\mathbf{R}_s\frac{\partial\Tilde{\mathbf{A}}^H}{\partial \zeta_n^{'}}\right\}\right\}.
\end{align}

\section{Numerical Results and Discussion}\label{Simulation}
In this section, we present simulation results demonstrating the performance of the proposed near-field multi-source localization methods. Unless otherwise stated, we consider an $11$-element ULA with half-wavelength inter-element spacing operating at the frequency $5$~GHz (i.e., $\lambda=6$~cm). The Fresnel region for this system setup is $[6.9\lambda,50\lambda]$. The results in all studied examples were obtained via averaging $K=500$ independent Monte Carlo trials. We have finally assumed that the MC is direction-dependent, in particular, the MC coefficients were generated using \cite{Elbir2017}'s method for each DOA.

\subsection{Performance Metrics and Investigated Estimators}
To evaluate the accuracy of the estimations for the DOAs, ranges, and direction-dependent MC coefficients, we have used the following root mean square error (RMSE) metrics:
\begin{align}\label{RMSE_DOA}
    &\text{RMSE}_\theta\triangleq\sqrt{\frac{1}{KN}\sum_{k=1}^{K}\sum_{n=1}^{N}(\hat{\theta}_{n,k}-\theta_{n})^2},\nonumber \\
    &\text{RMSE}_r\triangleq\sqrt{\frac{1}{KN}\sum_{k=1}^{K}\sum_{n=1}^{N}(\hat{r}_{n,k}-r_{n})^2},\nonumber \\
    &\text{RMSE}_c\triangleq\sqrt{\frac{1}{KNQ}\sum_{k=1}^{K}\sum_{n=1}^{N}\sum_{q=1}^{Q}(|\hat{c}_{n,k,q}-c_{n,q}|)^2}.
\end{align}
In addition, we used the signal-to-noise ratio (SNR) defined as the ratio of signal power to the noise power at the receiver.

For the proposed IMOP method, we have considered the value $0.01^\circ$ for the DOA convergence parameter $\varepsilon$. Furthermore, $\delta_\theta$ and $\delta_r$ were set to $0.1^\circ$ and $0.1\lambda$, respectively. The performance of the IMOP method was compared with the TSMNSL method, the 2D-MUSIC tailored for near-field localization~\cite{Huang1991}, and the two-stage rank reduction (TSRARE) method presented in~\cite{Xie2016}. The MC was assumed to be direction-independent and known (or estimated perfectly without any error) when implementing 2D-MUSIC and TSRARE methods. This idealistic assumption was made because these methods cannot estimate MC in a scenario where only near-field sources exist in the environment of interest. Furthermore, to implement 2D-MUSIC, we used the exact wavefront model. Note that, to the best of our knowledge, apart from our TSMNSL method, there is no other method for localizing near-field sources when considering direction-dependent MC.

\subsection{Results from Synthetic Simulations}
We have simulated in MATLAB version R2022b all afore-described near-field multi-source localization estimators for the following four examples.


\begin{figure}[!t]
\subfloat[]{\includegraphics[width=\columnwidth]{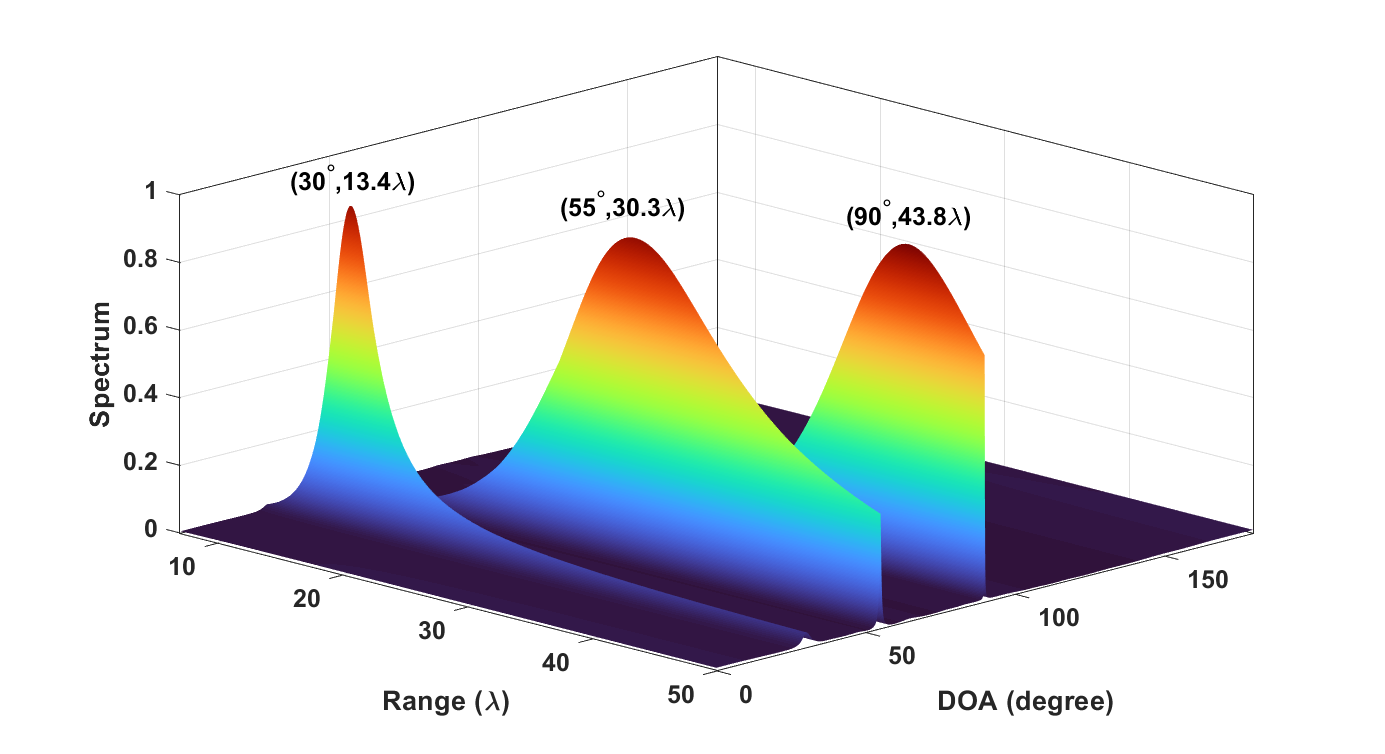}}%
\vfil
\subfloat[]{\includegraphics[width=\columnwidth]{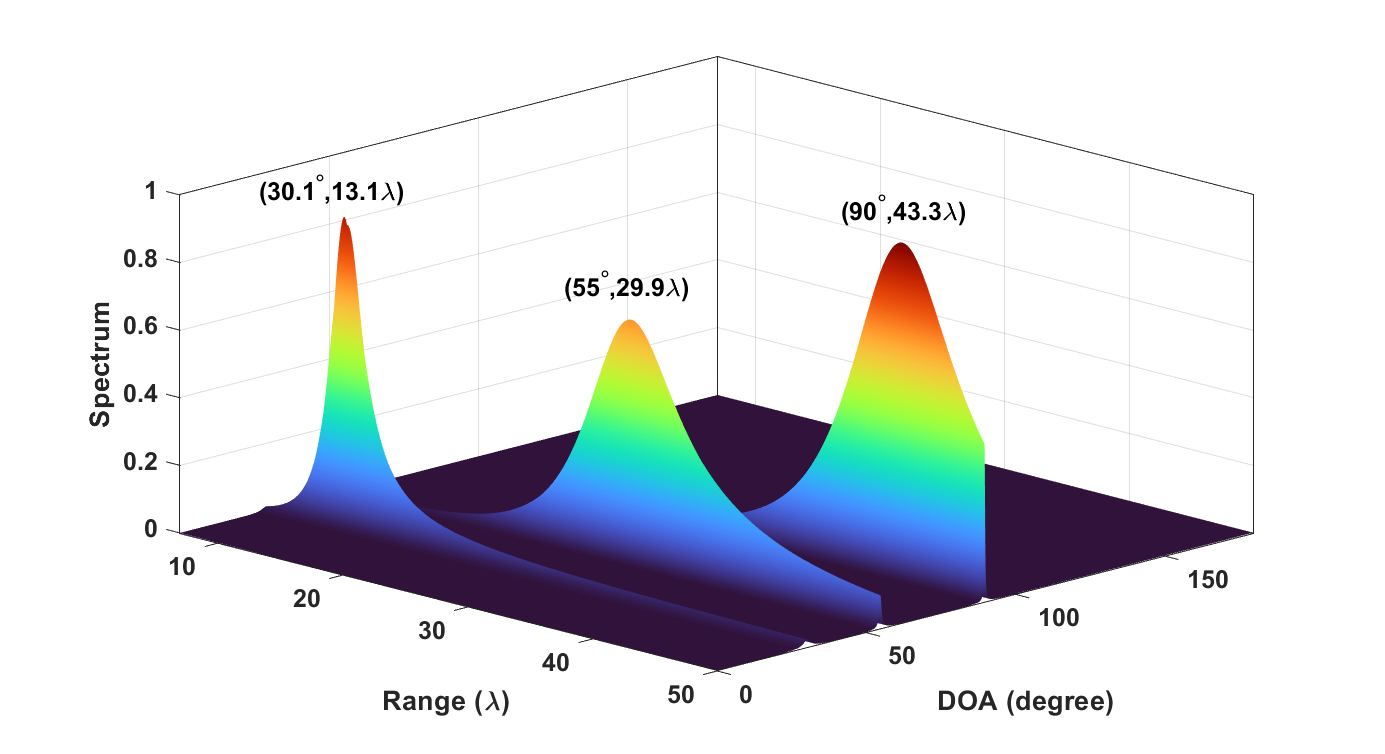}%
\label{RMSERangeVsSNR}}
\caption{The spectrum of (a) the proposed TSMNSL and (b) the 2D-MUSIC methods. The true positions of the three considered co-planar sources in Example 1 are $(30^\circ,13.3\lambda)$, $(55^\circ,30\lambda)$, and $(90^\circ,43.3\lambda)$.}
\label{Spectrum}
\end{figure}
\textit{Example 1 (Parameter Spectrum and Computational Load)}: There are three sources at the co-planar points $(30^\circ,13.3\lambda)$, $(55^\circ,30\lambda)$, and $(90^\circ,43.3\lambda)$. The number of snapshots $L$ and the SNR are fixed at $200$ and $10$~dB, respectively. 

\begin{figure}[!t]
\subfloat[]{\includegraphics[width=\columnwidth]{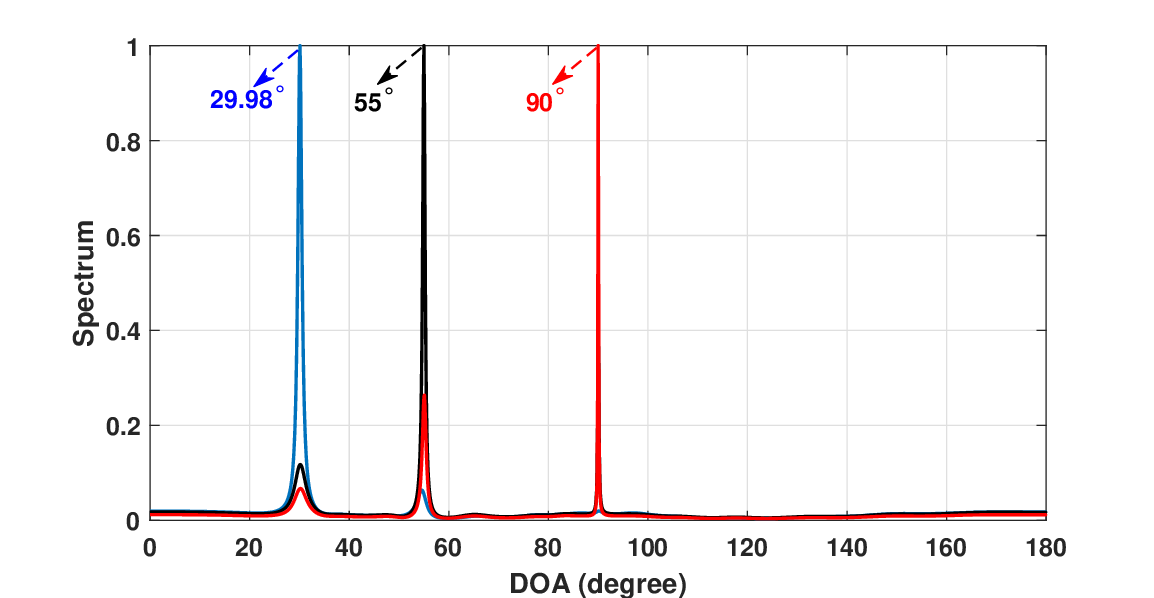}}%
\vfil
\subfloat[]{\includegraphics[width=\columnwidth]{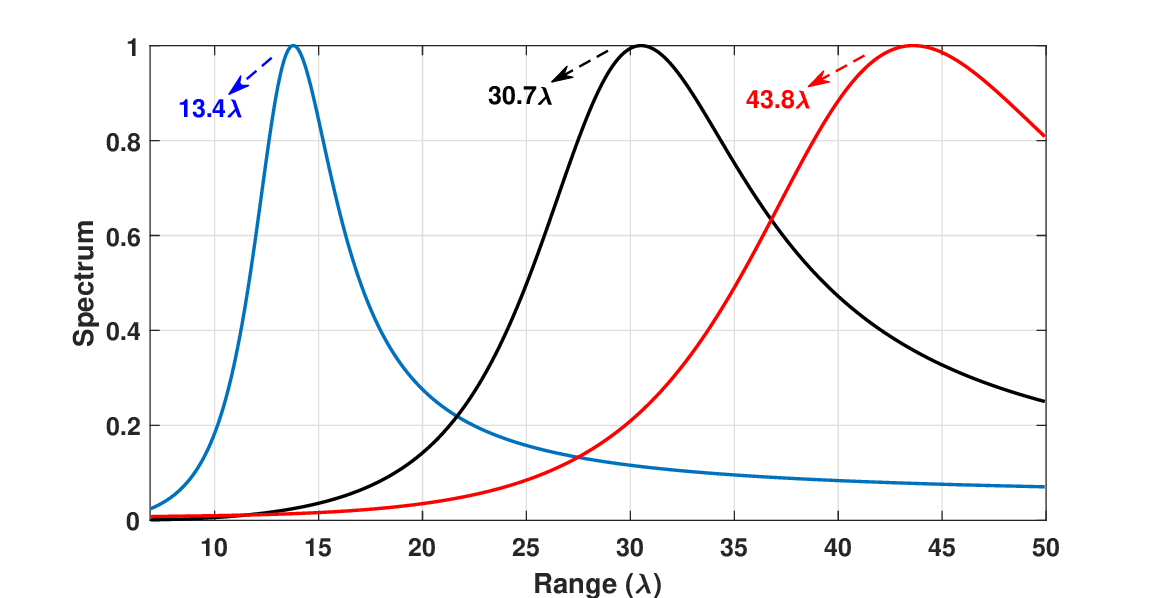}%
}
\caption{The spectrum of the proposed IMOP method for (a) DOA and (b) range estimation for the three sources of Example 1 considered also in Fig.~\ref{Spectrum}.}
\label{SpectrumSecondMethod}
\end{figure}

\begin{figure}[!t]
\subfloat[]{\includegraphics[width=\columnwidth]{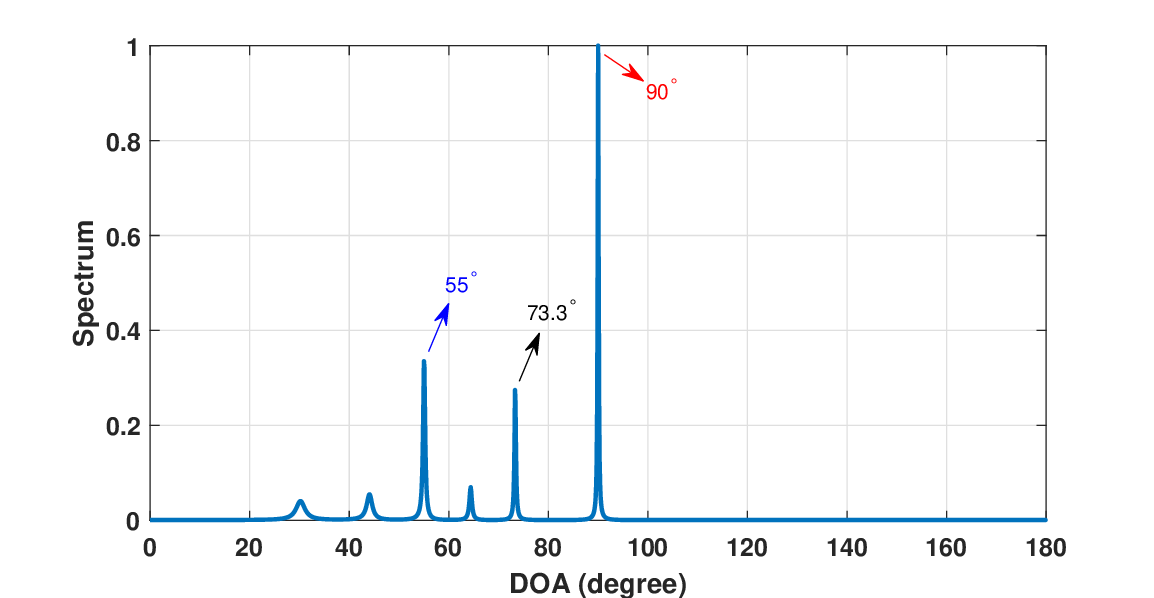}}
\vfil
\subfloat[]{\includegraphics[width=\columnwidth]{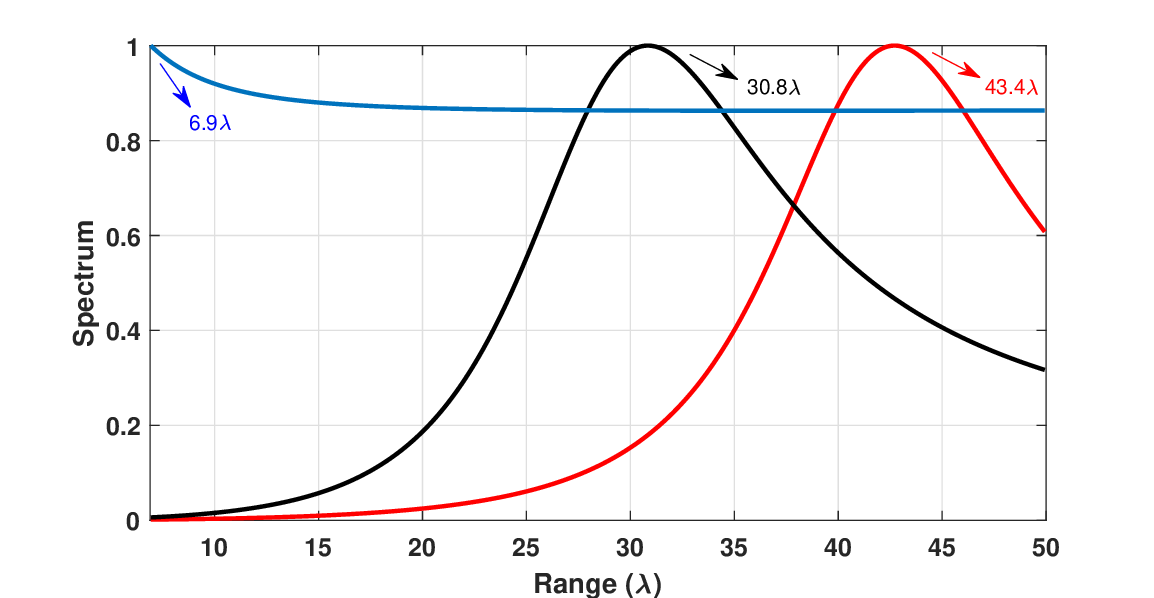}}
\caption{The spectrum of the TSRARE method for (a) DOA and (b) range estimation for the three sources of Example 1 also considered in Fig.~\ref{Spectrum}.}
\label{SpectrumTSRARE}
\end{figure}

\begin{figure}[!t]
    \centering
 \includegraphics[width=\columnwidth]{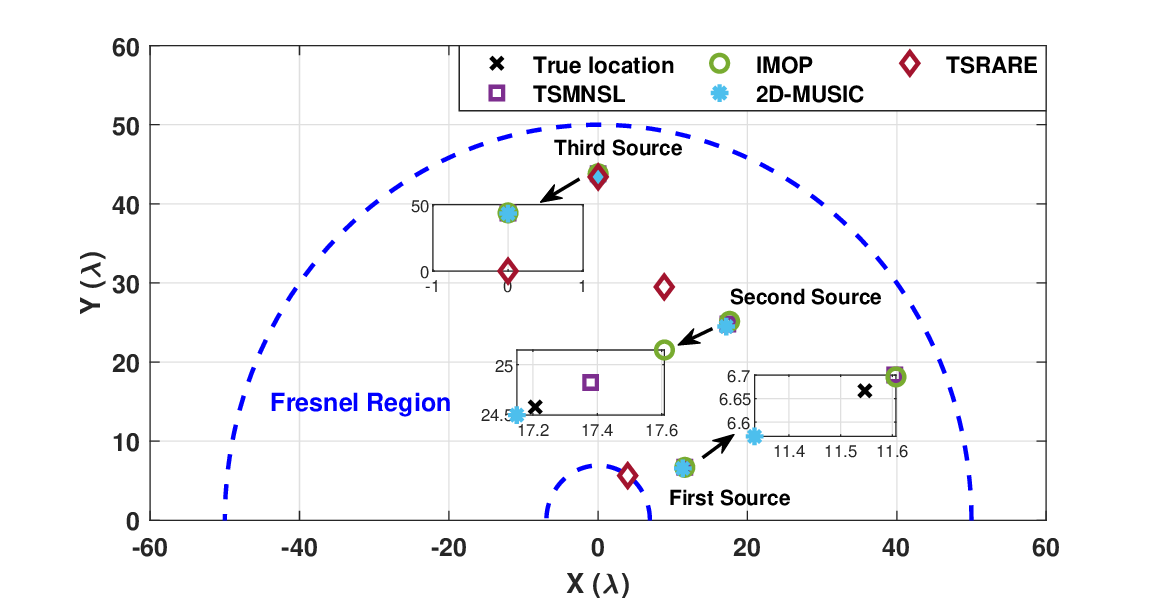}
    \caption{Illustration of the true and estimated locations of all three sources in Table~\ref{Tableexample1} for all considered near-field localization methods.}
    \label{LocationOfSources}
\end{figure}

Figures~\ref{Spectrum}(a) and~\ref{Spectrum}(b) as well as Fig.~\ref{SpectrumSecondMethod} depict the averaged 2D spatial spectrum (i.e., expression~\eqref{MUSIC_Spectrum} with known $\mathbf{c}_n$) for the proposed TSMNSL and 2D-MUSIC methods, as well as the averaged 1D spatial spectrum (i.e., \eqref{Estimated_DOA}) for the proposed IMOP method, respectively. As shown in all figures, three distinct peaks appear in the spectra indicating the location parameters of all three sources. The TSRARE spectrum for DOA and range are illustrated in Figs.~\ref{SpectrumTSRARE}(a) and~\ref{SpectrumTSRARE}(b), respectively. It can be observed that this localization method is unable to estimate the location of all sources. This is attributed to the fact that TSRARE is based on the approximate wavefront model and assumed direction-independent MC. The estimated DOAs and ranges using the 2D-MUSIC, TSRARE, TSMNSL, and IMOP methods are listed in Table~\ref{Tableexample1}. It is demonstrated that all methods, except TSRARE, can accurately the positions of all three sources. To better visualize the estimations, Fig.~\ref{LocationOfSources} shows the estimated positions of the sources in Cartesian coordinates. Evidently, the performance of the proposed IMOP method is almost similar to that of the TSMNSL and 2D-MUSIC methods. Recall that, to implement 2D-MUSIC, unlike the proposed TSMNSL and IMOP methods, we had to assume that the MC coefficients are known.


In Table~\ref{TableComputationalLoad}, we compare the computational load of all methods in Example 1 simulated in MATLAB running on an Intel(R) Core(TM) i7-1265U processor with $16$~GB of RAM. Recall that both the TSRARE and the proposed IMOP methods, unlike the two others, are based on 1D searches to estimate the location of the near-field sources. Hence, as it can be also observed in the table, their computational time is significantly less than that of the TSMNSL and 2D-MUSIC methods. In fact, the computation time of TSRARE is $7.5$ times less than that of the proposed IMOP method, since the former, unlike the latter, does not require TM construction. As depicted in Table~\ref{TableComputationalLoad}, $53.3\%$ of IMOP's total computation time is spent to calculate the required TMs. 
However, as shown in Table~\ref{Tableexample1}, TSRARE falls short in correctly estimating the positions of all three sources. It is also shown in Table~\ref{TableComputationalLoad} that the computation time of the 2D-MUSIC is $1.3$ times less than that of the TSMNSL method. This is because 2D-MUSIC does not form TMs, while $52.7\%$ of the computational load for the TSMNSL method is spent on this construction. Table~\ref{TableComputationalLoad} also demonstrates that the computation time of the proposed IMOP method is $42$ times less than that of the proposed TSMNSL method. This is because the former method is based on 1D searches, while the latter performs 2D searches.
\begin{table}[!t]
    \centering
    \caption{The estimated locations using the different methods.}
    \setlength{\tabcolsep}{2.2pt}
    \renewcommand{\arraystretch}{1.3}
    \begin{tabular}{|c|c|c|c|}
        \hline
        \textbf{True Positions} & $\mathbf{(30^\circ,13.3\lambda)}$ & $\mathbf{(55^\circ,30\lambda)}$ & $\mathbf{(90^\circ,43.3\lambda)}$ \\
        \hline\hline
        IMOP & $(29.98^\circ,13.4\lambda)$ & $(55.0^\circ,30.7\lambda)$ & $(90.0^\circ,43.8\lambda)$   \\ \hline
        TSMNSL & $(30.0^\circ,13.4\lambda)$ & $(55.0^\circ,30.3\lambda)$ & $(90.0^\circ,43.8\lambda)$  \\ \hline
        2D-MUSIC & $(30.1^\circ,13.1\lambda)$ & $(55.0^\circ,29.9\lambda)$ & $(90.0^\circ,43.3\lambda)$  \\ \hline
        TSRARE & $(55.0^\circ,6.9\lambda)$ & $(73.3^\circ,30.8\lambda)$ & $(90.0^\circ,43.4\lambda)$  \\
        \hline
    \end{tabular}\label{Tableexample1}
\end{table}

\begin{figure}[!t]
\centering
\subfloat[]{\includegraphics[width=\columnwidth]{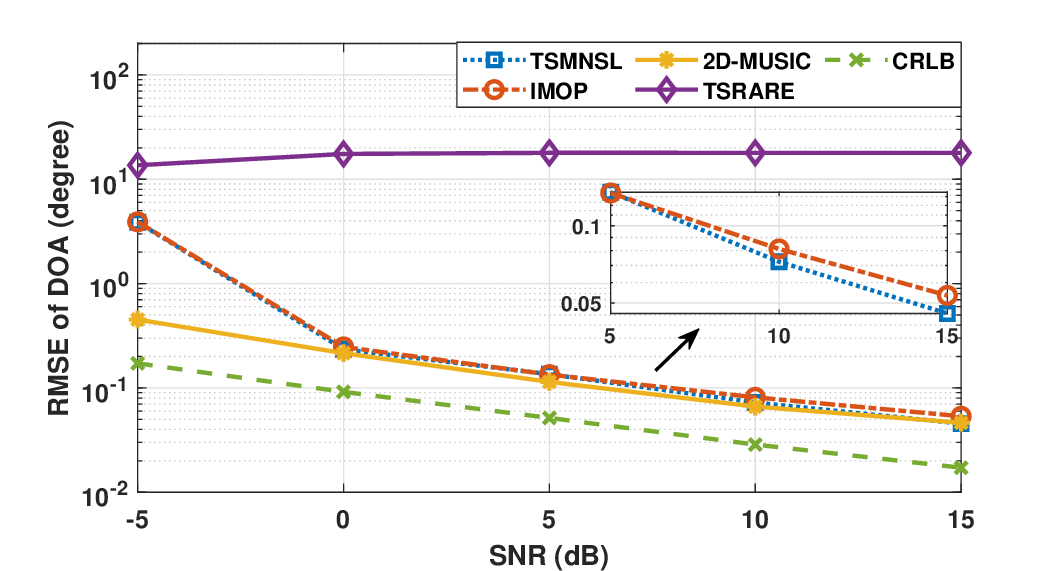}}
\vfil
\subfloat[]{\includegraphics[width=\columnwidth]{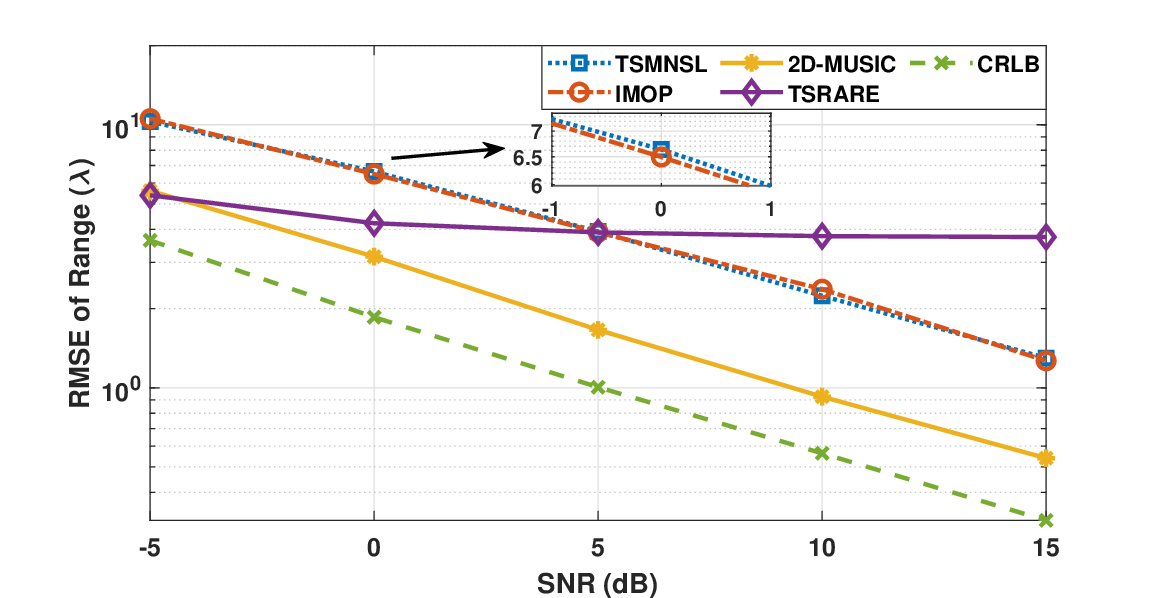}}
\vfil
\subfloat[]{\includegraphics[width=\columnwidth]{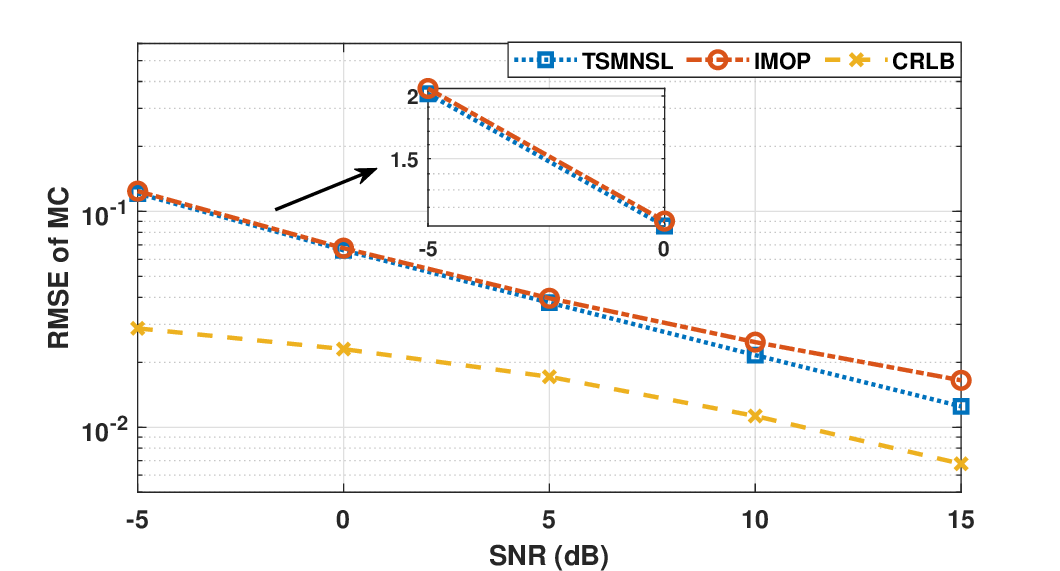}}
\caption{The RMSE of the estimated (a) DOAs, (b) ranges, and (c) MC versus the SNR in dB for the three sources of Example~1 considered also in Fig.~\ref{Spectrum}.}
\label{RMSEUnknownsVSSNR}
\end{figure}

\textit{Example 2 (Position Estimation Performance versus SNR)}: In Figs.~\ref{RMSEUnknownsVSSNR}(a), \ref{RMSEUnknownsVSSNR}(b), and \ref{RMSEUnknownsVSSNR}(c), we consider the same setup with Example 1 and plot the averaged RMSE of the estimated DOAs, ranges, and MC with respect to the SNR that varies from $-5$~dB to $15$~dB with a $5$~dB interval. As expected for all methods, the estimation performance improves with increasing SNR values.
\begin{table}[!t]
    \centering
    \caption{The computational load of the different methods.}
    \setlength{\tabcolsep}{2.2pt}
    \renewcommand{\arraystretch}{1.2}
    \resizebox{\columnwidth}{!}{%
    \begin{tabular}{|c|c|c|}
        \hline
        \textbf{Method} & \textbf{TMs construction time (sec)} & \textbf{Total computation time (sec)} \\
        \hline\hline
         IMOP & $0.08$& $0.15$\\ \hline
        TSMNSL & $3.31$& $6.28$\\ \hline
        2D-MUSIC & - & $4.78$\\ \hline
        TSRARE & - & $0.02$\\
        \hline
    \end{tabular}}
    \label{TableComputationalLoad}
\end{table}
It can be observed from Fig.~\ref{RMSEUnknownsVSSNR}(a) that the DOA estimation performance with the proposed IMOP and TSMNSL methods is very close to each other in low SNRs, specifically for SNR less than or equal to $5$~dB, while the computational complexity of the IMOP method is considerably less than TSMNSL's one. It is also shown that the TSMNSL method performs slightly better than IMOP at high SNRs, in particular, for SNRs higher than $5$~dB. For instance, for the SNR value of $15$~ dB, the RMSE of the estimated DOA using the TSMNSL and IMOP methods is $0.046^\circ$ and $0.054^\circ$, respectively. Figure~\ref{RMSEUnknownsVSSNR}(a) also illustrates that, for SNRs less than $0$~dB, the RMSE of the DOAs estimated by the 2D-MUSIC method is significantly less than that with other methods. This can be due to the fact that 2D-MUSIC estimates DOAs while assuming the MC is known (this a-priori information may not be available in practical applications). For example, the RMSE of DOA estimation with 2D-MUSIC, TSMNSL, IMOP, and TSRARE at the SNR value of $-5$ dB is $0.45^\circ$, $3.88^\circ$, $3.90^\circ$, and $13.6^\circ$, respectively. However, for SNR values equal to or greater than $0$~dB, the performance of the proposed TSMNSL and IMOP methods is close to the 2D-MUSIC performance. This is because the effect of the MC on the performance of IMOP and TSMNSL methods in DOA estimation at high SNRs is small. For the SNR at $10$~dB, the RMSE of DOA estimation using TSMNSL, IMOP, and 2D-MUSIC is $0.072^\circ$, $0.081^\circ$, and $0.066^\circ$, respectively. It is finally shown in Fig.~\ref{RMSEUnknownsVSSNR}(a) that the RMSE of the estimated DOA using TSRARE is significantly higher than that with the other methods for all SNR values. This is due to the fact that TSRARE is based on the approximation wavefront model.

Figure~\ref{RMSEUnknownsVSSNR}(b) depicts the RMSE of the range estimation for all simulated near-field localization methods. Clearly, 2D-MUSIC is the best method across all SNR values except at $-5$~dB. At this value, TSRARE is the best method yielding the estimation $5.39\lambda$, whereas 2D-MUSIC, TSMNSL, and IMOP provide the range estimations $5.61\lambda$, $10.29\lambda$, and $10.55\lambda$, respectively. At SNR of $0$~dB, the RMSE of the estimated range using 2D-MUSIC, TSRARE, IMOP, and TSMNSL is $3.15\lambda$, $4.22\lambda$, $6.50\lambda$, and $6.64\lambda$, respectively. For SNRs larger than $5$~dB, the performance of TSMNSL and IMOP methods is better than that of TSRARE. It is finally shown that 2D-MUSIC yields the closest performance to the CRLB of the range estimation.

The RMSE of the MC estimation for both methods, TSMNSL and IMOP, is depicted in Fig.~\ref{RMSEUnknownsVSSNR}(c) together with the respective CRLB. It can be seen that TSMNSL outperforms IMOP across all considered SNR values. In fact, the difference between these two methods is very small. For example, at an SNR of $0$~dB, the RMSE of the estimated MC using the TSMNSL and IMOP is $0.066$ and $0.068$, respectively. When the SNR increases from $5$~dB to $15$~dB, the difference in the estimation performance also increases. For example, at the SNR value of $15$~dB, the RMSE performance with the TSMNSL method is $0.013$, while with IMOP is $0.017$.

\begin{figure}[!t]
\centering
\subfloat[]{\includegraphics[width=\columnwidth]{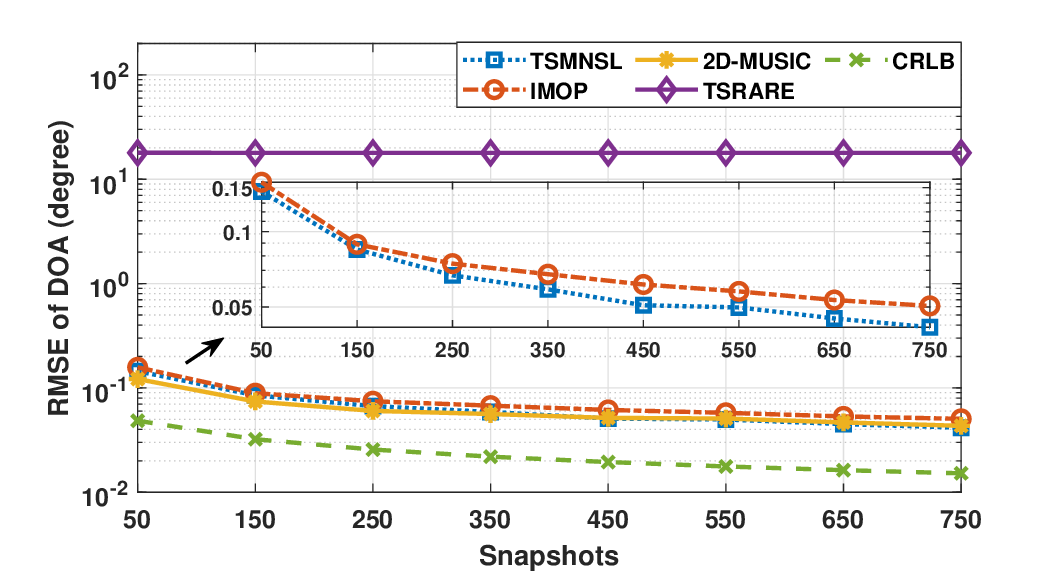}}
\vfil
\subfloat[]{\includegraphics[width=\columnwidth]{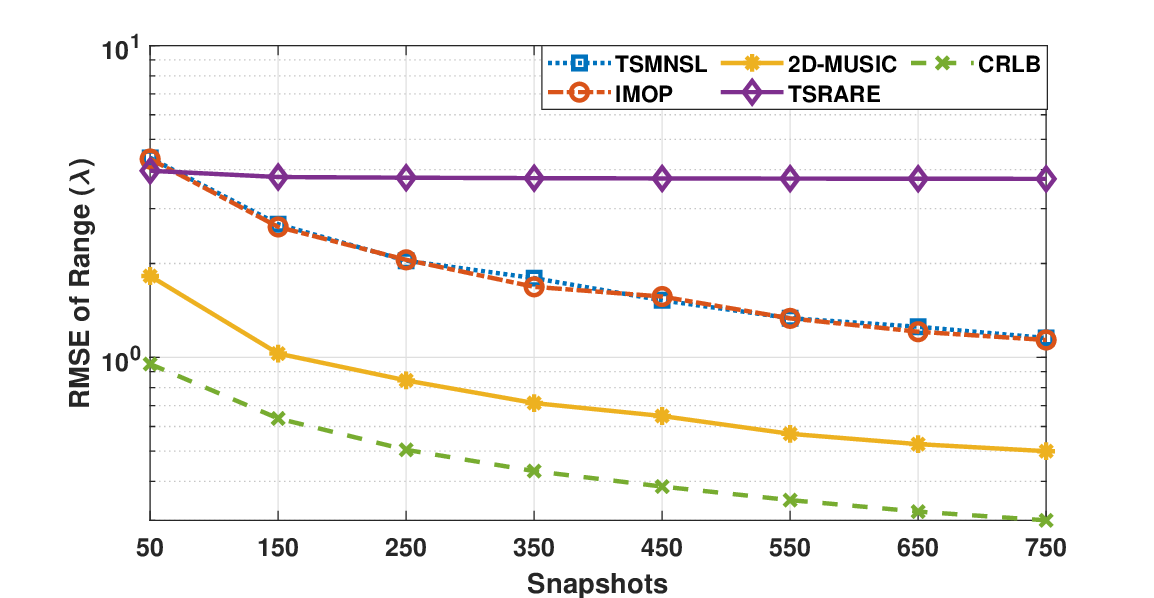}}
\vfil
\subfloat[]{\includegraphics[width=\columnwidth]{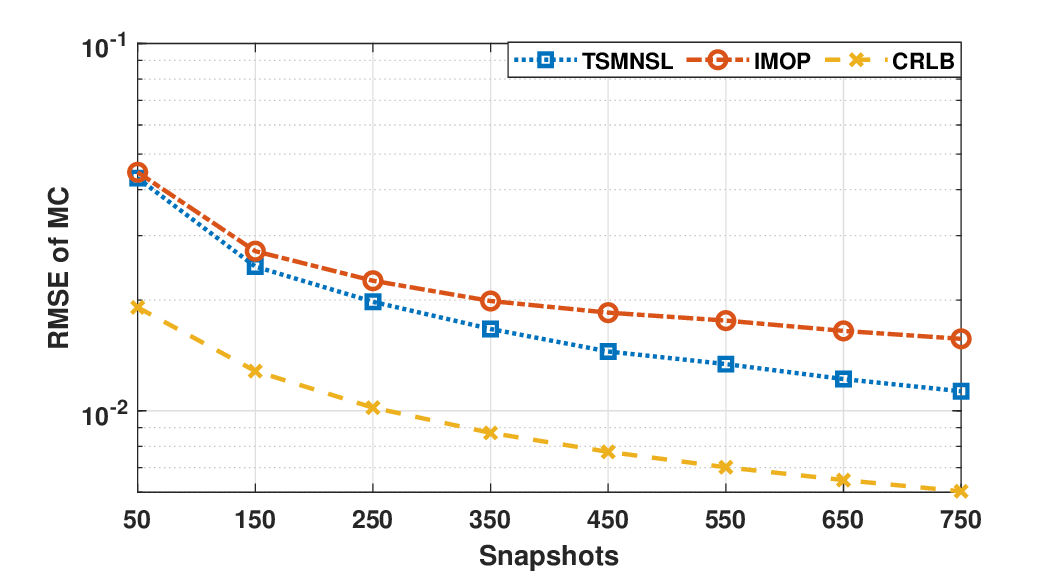}}
\caption{The RMSE of the estimated (a) DOAs, (b) ranges, and (c) MC versus the number of snapshots $L$ for the three sources of Example~1 considered also in Fig.~\ref{Spectrum}.}
\label{RMSEUnknownsVSSnap}
\end{figure}

\textit{Example 3 (Position Estimation Performance versus Number of Snapshots)}:
We have set the SNR at $10$~dB in Figs.~\ref{RMSEUnknownsVSSnap}(a), \ref{RMSEUnknownsVSSnap}(b), and \ref{RMSEUnknownsVSSnap}(c) for the same setup with Example 1 and varied the number of snapshots $L$ from $50$ to $750$ with a step of $100$. It can be seen from Fig.~\ref{RMSEUnknownsVSSnap}(a) that 2D-MUSIC outperforms all other methods in estimating DOA for every tested value for $L$. This happens because we have implemented 2D-MUSIC with the exact wavefront model considering that the MC is known. To better compare the performance with the proposed TSMNSL and IMOP methods, this figure also includes an inset zoomed-in box. It is shown that TSMNSL outperforms the IMOP method for all numbers of snapshots. For example, for $L=350$, the RMSE of the estimated DOA using TSMNSL and IMOP method is $0.059^\circ$ and $0.068^\circ$, respectively. Note that the difference between the RMSEs of TSMNSL and IMOP increases as $L$ increases from $150$ to $750$. 

Figure~\ref{RMSEUnknownsVSSnap}(b) shows that 2D-MUSIC outperforms all other methods in the range estimation for all $L$ values. Recall though that 2D-MUSIC assumes that MC is known, which may not be the case in the practical scenarios. The figure also demonstrates that, for $L=50$, the RMSE of the range estimated using 2D-MUSIC, TSMNSL, IMOP, and TSRARE is $1.82\lambda$, $4.36\lambda$, $4.33\lambda$ and $3.97\lambda$, respectively. For $L$ values larger than $50$, the performance of TSRARE compared to TSMNSL and IMOP degrades. Additionally, for $L=750$, the RMSE of the range estimated using 2D-MUSIC, TSMNSL, IMOP, and TSRARE is $0.5\lambda$, $1.15\lambda$, $1.14\lambda$, and $3.74\lambda$, respectively. Finally, Fig.~\ref{RMSEUnknownsVSSnap}(c) demonstrates that the CRLB of the MC estimation and the respective RMSE using the proposed TSMNSL and IMOP methods decrease with increasing $L$ values. It is also shown that the performance with the IMOP method is very close to that with TSMNSL for small $L$ values. For example, for $L=50$, the RMSEs of the estimated MC using the TSMNSL and IMOP methods are $0.043$ and $0.045$, respectively. Although the MC estimation difference between TSMNSL and IMOP increases as $L$ increases, this is insignificant for the simulated $L$ range. For example, for $L=750$, the RMSEs of the MC estimation with the proposed TSMNSL and IMOP methods are $0.011$ and $0.016$, respectively.
\begin{figure}[!t]
\subfloat[]{\includegraphics[width=\columnwidth]{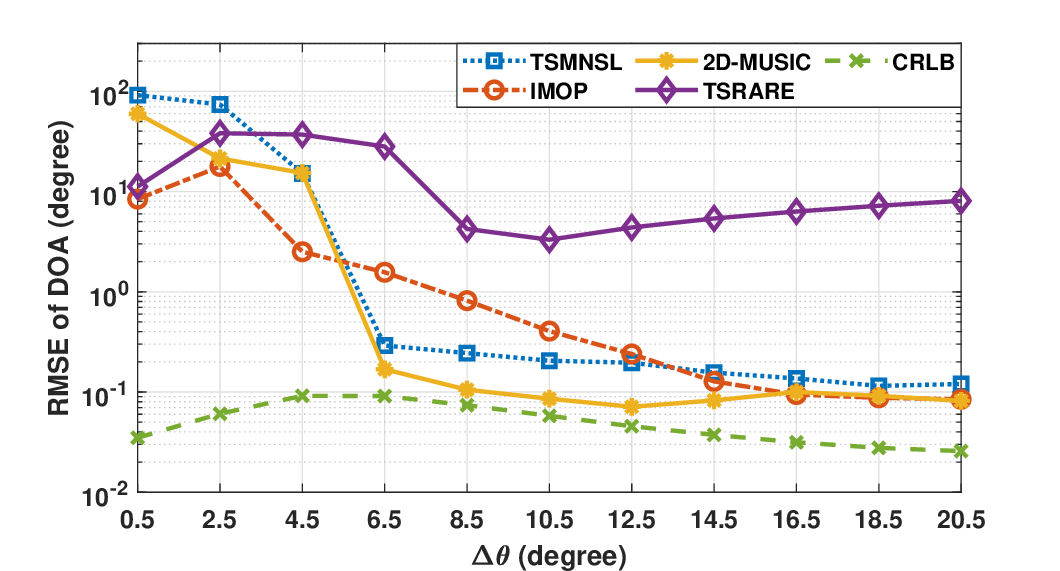}}
\vfil
\subfloat[]{\includegraphics[width=\columnwidth]{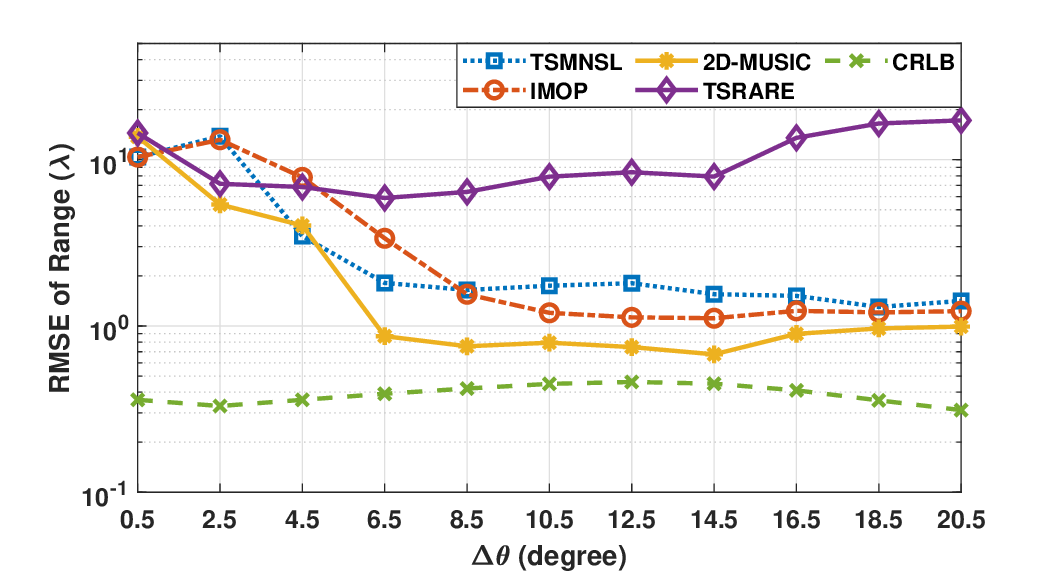}}
\caption{The RMSE of the estimated (a) DOAs and (b) ranges versus the angular separation $\Delta\theta$ for the co-planar sources $(30^\circ,13.3\lambda)$ and $(30^\circ+\Delta\theta,30\lambda)$ in Example 4.}
\label{RMSEUnknownVsAngularSep}
\end{figure}


\textit{Example 4 (Position Estimation Performance versus Source Angular Separation)}: Consider two sources located at the co-planar positions $(30^\circ,13.3\lambda)$ and $(30^\circ+\Delta\theta,30\lambda)$, where $\Delta\theta$ represents their angular separation, which is assumed to vary from $0.5^\circ$ to $20.5^\circ$ at an angular interval of $2^\circ$. In Figs.~\ref{RMSEUnknownVsAngularSep}(a) and \ref{RMSEUnknownVsAngularSep}(b), we have considered the values $15$~dB and $200$ for the SNR and the number of snapshots $L$, respectively, and depict the RMSEs of the estimated DOAs and ranges versus $\Delta\theta$.

It can be observed from Fig.~\ref{RMSEUnknownVsAngularSep}(a) that the proposed IMOP method outperforms all other methods for $\Delta\theta$ ranging from $0.5^\circ$ to $4.5^\circ$. As $\Delta\theta$ increases from $6.5^\circ$ to $12.5^\circ$, the RMSEs of DOA estimation using 2D-MUSIC and TSMNSL become lower than those obtained using IMOP and TSRARE. For example, for $\Delta\theta=8.5^\circ$, the RMSE values for DOA estimation using the 2D-MUSIC, TSMNSL, IMOP, and TSRARE are $0.11^\circ$, $0.24^\circ$, $0.82^\circ$, and $4.24^\circ$, respectively. In the case of high $\Delta\theta$, i.e., from $16.5^\circ$ to $20.5^\circ$, the proposed IMOP method performs similarly to 2D-MUSIC and better than the other two methods. It can be finally observed from Fig.~\ref{RMSEUnknownVsAngularSep}(b) shows that, as $\Delta\theta$ increases, the RMSE of range estimation with IMOP becomes closer to that with 2D-MUSIC and superior to the other two methods. For example, when $\Delta\theta=20.5^\circ$, the RMSEs of the estimated range using the 2D-MUSIC, IMOP, and TSMNSL are $\lambda$, $1.2\lambda$, and $1.4\lambda$, respectively.

\begin{figure}[!t]
\subfloat[]{\includegraphics[width=\columnwidth]{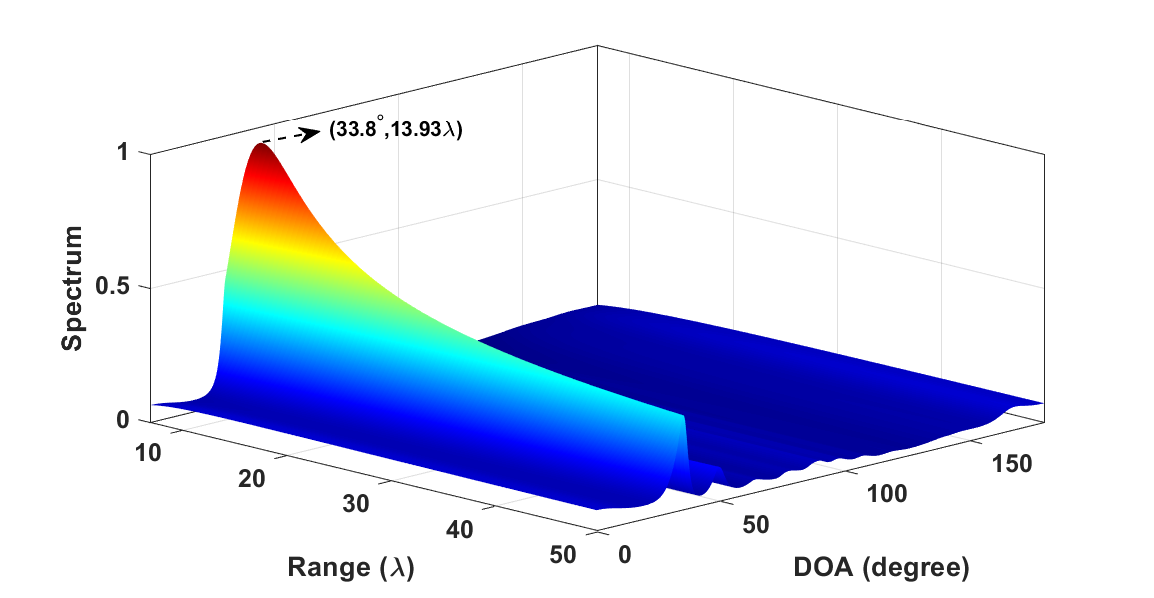}}%
\vfil
\subfloat[]{\includegraphics[width=\columnwidth]{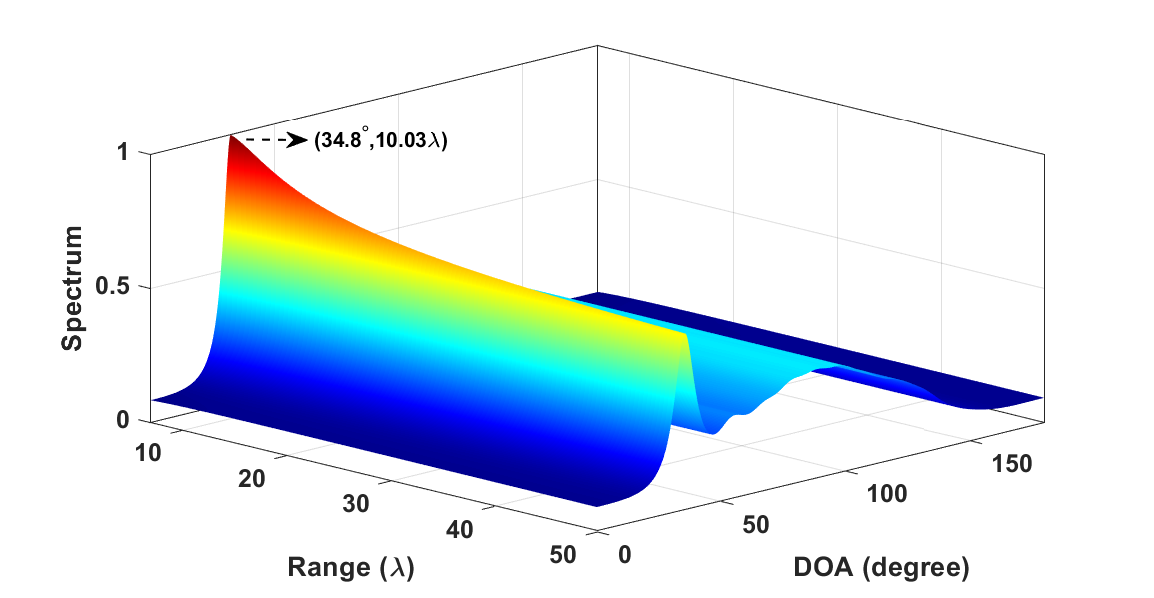}}
\caption{The spectrum obtained from full-wave electromagnetic simulations with the (a) proposed TSMNSL and (b) 2D-MUSIC methods, considering a single source located at the position $(35^\circ,13.3\lambda)$.}
\label{SpectrumCST1}
\end{figure}
\subsection{Results from Full-Wave Electromagnetic Simulations}\label{DataGeneratedUsingCST}
We now present full-wave electromagnetic simulations using CST Microwave Studio~\cite{CSTStudio} for the considered near-field localization methods. These simulations allow us to verify the efficacy of the proposed techniques in setups being closer to realistic scenarios and, thus, making them more applicable to practical near-field applications. It is critical to verify the techniques using non-isotropic antennas, (at least omnidirectional) to study the EM interaction and MC dependence on the element's relative orientation and positioning within the array~\cite{Larmour2024}. We considered a ULA-based receiver consisting of $11$ half-wavelength dipole antennas operating at the frequency $5$~GHz tasked to localize a single source with a DOA and range of $35^\circ$ and $13.3\lambda$, respectively. All localization algorithms 2D-MUSIC, TSRARE, TSMNSL, and IMOP were implemented in MATLAB and we have imported directly from CST full-wave electromagnetic simulation data. 

\begin{figure}[!t]
\subfloat[]{\includegraphics[width=\columnwidth]{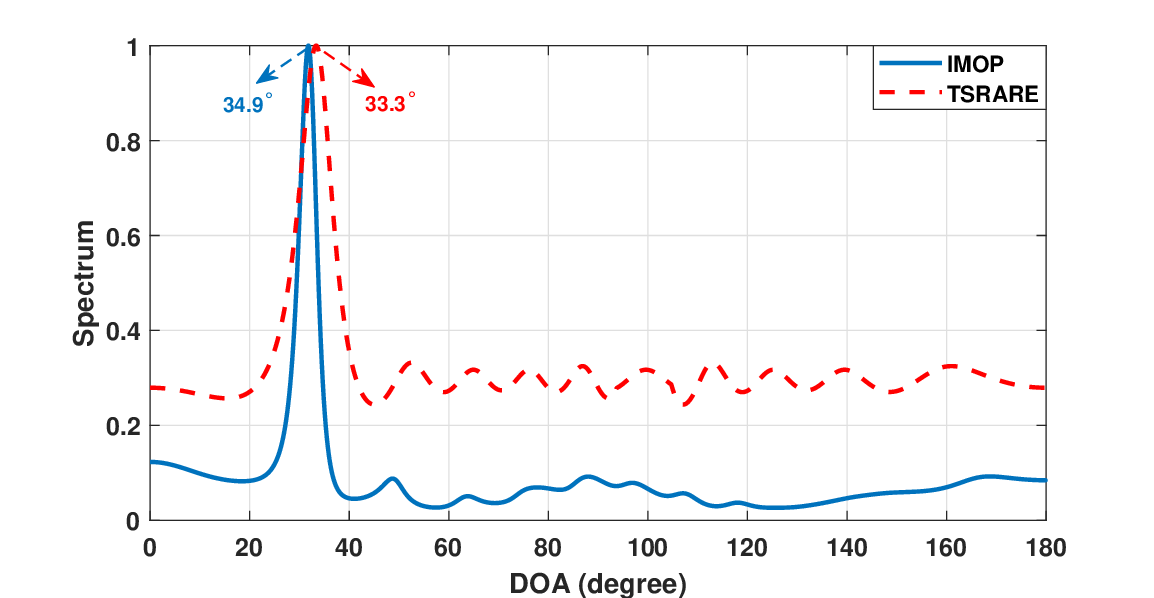}}
\vfil
\subfloat[]{\includegraphics[width=\columnwidth]{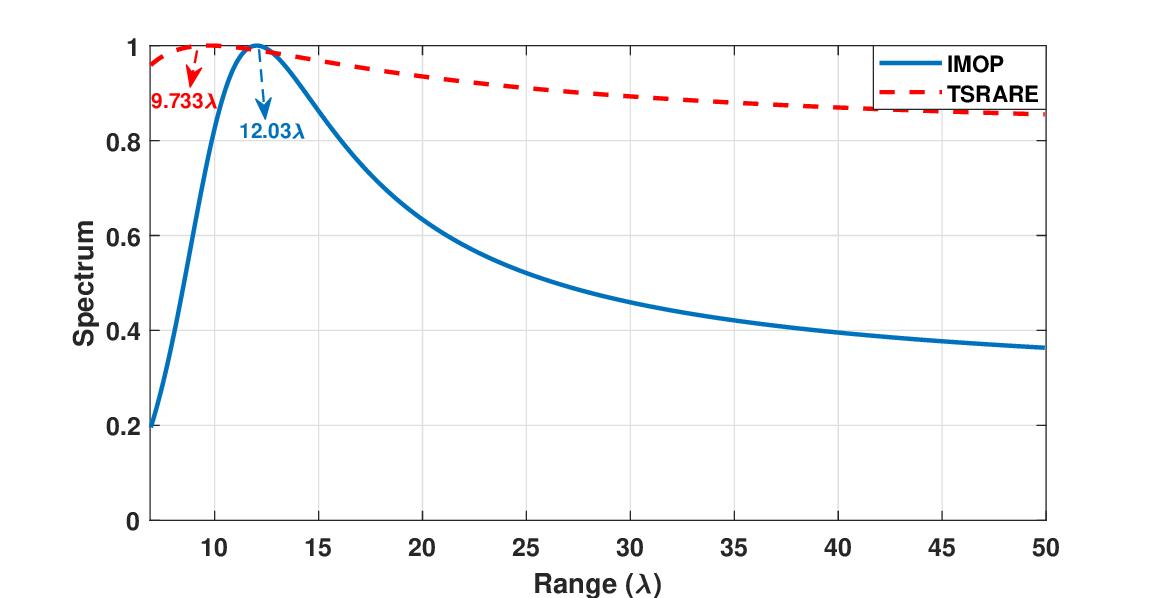}}
\caption{The spectrum obtained from full-wave electromagnetic simulations with the (a) proposed IMOP and (b) TSRARE methods for the same single source with Fig.~\ref{SpectrumCST1}.}
\label{SpectrumCST2}
\end{figure}

In Figs.~\ref{SpectrumCST1}(a) and \ref{SpectrumCST1}(b), the estimated 2D spatial spectra with the TSMNSL and 2D-MUSIC methods are illustrated, respectively. The sharp peak in these figures shows that both methods can estimate the location of the single source. In Fig.~\ref{SpectrumCST1}, we can observe that the peak of 2D-MUSIC is sharper than that from the TSMNSL method. Recall that, for the implementation of the 2D-MUSIC, we have assumed that the MC is perfectly known. In addition, Figs.~\ref{SpectrumCST2}(a) and~\ref{SpectrumCST2}(b) depict respectively the DOA and range spatial spectra of the IMOP and TSRARE methods. It can be seen that the proposed IMOP method outperforms TSRARE, indicating that it is more compatible with the full-wave-based realistic scenario.

\begin{figure}[!t]
    \centering
 \includegraphics[width=\columnwidth]{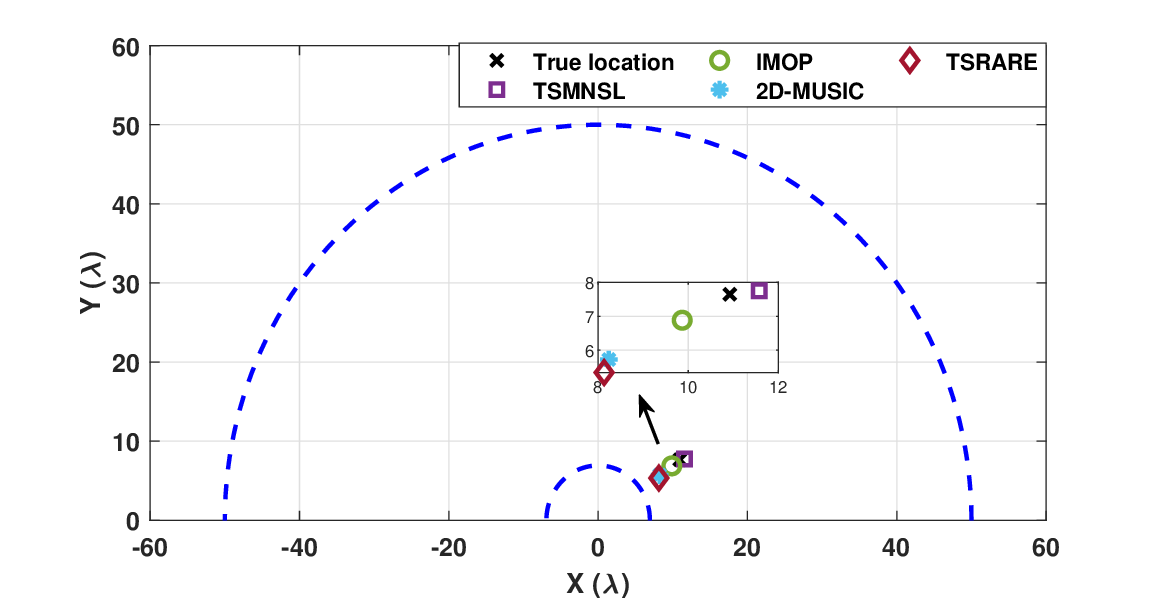}
    \caption{Illustration of the true and estimated locations of the single source in Table~\ref{TableCST} for all considered near-field localization methods.}
    \label{LocationOfSourceCST}
\end{figure}

Table~\ref{TableCST} includes the values of the estimated locations of the single source using the TSMNSL, IMOP, 2D-MUSIC, and TSRARE methods. It is shown that the best DOA estimation is provided by the proposed IMOP method, while the best range estimation is given by the proposed TSMNSL method. This table is in agreement with Fig.~\ref{LocationOfSourceCST}, demonstrating that the TSMNSL method can best estimate the source's location at the cost of the largest computational complexity. Interestingly, the performance with the proposed IMOP method is close to TSMNSL with significantly lower computational complexity. 
\begin{table}[!t]
    \centering
    \caption{The estimated locations of the source retrieved using different methods.}
    \setlength{\tabcolsep}{1.5pt}
    \renewcommand{\arraystretch}{1.8}
    \resizebox{\columnwidth}{!}{%
    \begin{tabular}{|c|c|c|c|c|}
        \hline
        \textbf{Ground Truth} & \textbf{IMOP} & \textbf{TSMNSL} & \textbf{2D-MUSIC} & \textbf{TSRARE} \\
        \hline
         $(35^\circ,13.3\lambda)$ & 
         $(34.9^\circ,12.03\lambda)$ & $(33.8^\circ,13.93\lambda)$ & $(34.8^\circ,10.03\lambda)$ & $(33.3^\circ,9.73\lambda)$\\
        \hline
    \end{tabular}%
    }
    \label{TableCST}
\end{table}

\section{Conclusions}\label{Conclusion}
In this paper, we presented the IMOP method to estimate the DOA and range of sources lying in the near-field region of a ULA-equipped receiver, taking into account the direction-dependent MC between the elements of the ULA. This iterative IMOP method first estimates the DOAs and ranges of multiple sources via 1D searches and then deploys them to estimate the direction-dependent MC, while the TSMNSL method performs 2D searches to perform these estimations. In addition, IMOP is equipped with an oblique projection operator to isolate the components of one near-field source from those of other sources. Our extensive simulation results showcased that the performance of the lower complexity IMOP method is similar to that of the TSMNSL method. For example, for the localization of three near-field sources with an $11$-element ULA receiver, IMOP and TSMNSL required $0.15$ and $6.28$ seconds of estimation computation time, respectively. This computation improvement is attributed to the efficient replacement of 2D searches from 1D ones for the estimations of DOAs and ranges of multiple sources.  

\bibliographystyle{IEEEtran}
\bibliography{IEEEabrv,BibliographyForMutualCouplingJournal}

\begin{thebibliography}{10}
\providecommand{\url}[1]{#1}
\csname url@samestyle\endcsname
\providecommand{\newblock}{\relax}
\providecommand{\bibinfo}[2]{#2}
\providecommand{\BIBentrySTDinterwordspacing}{\spaceskip=0pt\relax}
\providecommand{\BIBentryALTinterwordstretchfactor}{4}
\providecommand{\BIBentryALTinterwordspacing}{\spaceskip=\fontdimen2\font plus
\BIBentryALTinterwordstretchfactor\fontdimen3\font minus \fontdimen4\font\relax}
\providecommand{\BIBforeignlanguage}[2]{{%
\expandafter\ifx\csname l@#1\endcsname\relax
\typeout{** WARNING: IEEEtran.bst: No hyphenation pattern has been}%
\typeout{** loaded for the language `#1'. Using the pattern for}%
\typeout{** the default language instead.}%
\else
\language=\csname l@#1\endcsname
\fi
#2}}
\providecommand{\BIBdecl}{\relax}
\BIBdecl

\bibitem{Tao2022}
Q.~Tao, Z.~Hu, Z.~Zhou, H.~Xiao, and J.~Zhang, ``Seqpolar: Sequence matching of polarized lidar map with hmm for intelligent vehicle localization,'' \emph{IEEE Transactions on Vehicular Technology}, vol.~71, no.~7, pp. 7071--7083, 2022.

\bibitem{Molaei2022}
A.~M. Molaei, P.~del Hougne, V.~Fusco, and O.~Yurduseven, ``Efficient joint estimation of {DOA}, range and reflectivity in near-field by using mixed-order statistics and a symmetric {MIMO} array,'' \emph{{IEEE} Transactions on Vehicular Technology}, vol.~71, no.~3, pp. 2824--2842, 2022.

\bibitem{Luan2022}
M.~Luan, B.~Wang, Y.~Zhao, Z.~Feng, and F.~Hu, ``Phase design and near-field target localization for ris-assisted regional localization system,'' \emph{IEEE Transactions on Vehicular Technology}, vol.~71, no.~2, pp. 1766--1777, 2022.

\bibitem{Zhang2023}
Q.~Zhang, W.~Li, B.~Yang, and S.~Li, ``An auxiliary source-based near field source localization method with sensor position error,'' \emph{Signal Processing}, vol. 209, p. 109039, aug 2023.

\bibitem{Position_aided}
G.~C. Alexandropoulos, ``Position aided beam alignment for millimeter wave backhaul systems with large phased arrays,'' in \emph{Proc. IEEE Workshop Computational Adv. Multi-Sensor Adaptive Process.}, Cura\c{c}ao, Dutch Antilles, Dec. 2017.

\bibitem{Jang2024}
S.~Jang and C.~Lee, ``{DNN}-driven single-snapshot near-field localization for hybrid beamforming systems,'' \emph{IEEE Transactions on Vehicular Technology}, vol.~73, no.~7, pp. 10\,799--10\,804, 2024.

\bibitem{AVW22a}
G.~C. Alexandropoulos, I.~Vinieratou, and H.~Wymeersch, ``Localization via multiple reconfigurable intelligent surfaces equipped with single receive {RF} chains,'' \emph{IEEE Wireless Commun. Lett.}, vol.~11, no.~5, pp. 1072--1076, May 2022.

\bibitem{R_RIS}
J.~He, A.~Fakhreddine, C.~Vanwynsberghe, H.~Wymeersch, and G.~C. Alexandropoulos, ``{3D} localization with a single partially-connected receiving {RIS}: Positioning error analysis and algorithmic design,'' \emph{IEEE Transactions on Vehicular Technology}, vol.~10, no.~10, pp. 13\,190--13\,202, 2023.

\bibitem{Ghazalian2024}
R.~Ghazalian, G.~C. Alexandropoulos, G.~Seco-Granados, H.~Wymeersch, and R.~Jantti, ``Joint 3d user and 6d hybrid reconfigurable intelligent surface localization,'' \emph{IEEE Transactions on Vehicular Technology}, pp. 1--15, 2024.

\bibitem{Ghazalian2024b}
R.~Ghazalian, H.~Chen, G.~C. Alexandropoulos, G.~Seco-Granados, H.~Wymeersch, and R.~Jäntti, ``Joint user localization and location calibration of a hybrid reconfigurable intelligent surface,'' \emph{IEEE Transactions on Vehicular Technology}, vol.~73, no.~1, pp. 1435--1440, 2024.

\bibitem{Guanghui2020}
C.~Guanghui, Z.~Xiaoping, J.~Shuang, Y.~Anning, and L.~Qi, ``High accuracy near-field localization algorithm at low {SNR} using fourth-order cumulant,'' \emph{{IEEE} Communications Letters}, vol.~24, no.~3, pp. 553--557, mar 2020.

\bibitem{AbuShaban2021}
Z.~Abu-Shaban, K.~Keykhosravi, M.~F. Keskin, G.~C. Alexandropoulos, G.~Seco-Granados, and H.~Wymeersch, ``Near-field localization with a reconfigurable intelligent surface acting as lens,'' in \emph{Proc. IEEE Int. Conf. Commun.}, Montreal, Canada, Jun. 2021.

\bibitem{Li2021}
J.~Li, Y.~Wang, Z.~Ren, X.~Gu, M.~Yin, and Z.~Wu, ``{DOA} and range estimation using a uniform linear antenna array without a priori knowledge of the source number,'' \emph{{IEEE} Transactions on Antennas and Propagation}, vol.~69, no.~5, pp. 2929--2939, may 2021.

\bibitem{Cheng2022}
C.~Cheng, S.~Liu, H.~Wu, and Y.~Zhang, ``An efficient maximum-likelihood-like algorithm for near-field coherent source localization,'' \emph{{IEEE} Transactions on Antennas and Propagation}, vol.~70, no.~7, pp. 6111--6116, jul 2022.

\bibitem{RIS_near_field}
M.~Rahal, B.~Denis, K.~Keykhosravi, M.~F. Keskin, B.~Uguen, G.~C. Alexandropoulos, and H.~Wymeersch, ``Performance of {RIS}-aided nearfield localization under beams approximation from real hardware characterization,'' \emph{EURASIP J. Wireless Commun. and Netw.}, vol. 2023, no.~86, pp. 1--23, Aug. 2023.

\bibitem{gavras2023_hmimo_ISAC}
I.~Gavras, M.~A. Islam, B.~Smida, and G.~C. Alexandropoulos, ``Full duplex holographic {MIMO} for near-field integrated sensing and communications,'' in \emph{Proc. European Signal Process. Conf.}, Helsinki, Finland, Sep. 2023.

\bibitem{gavras2024dma_1bit}
I.~Gavras, I.~Atzeni, and G.~C. Alexandropoulos, ``Near-field localization with 1-bit quantized hybrid {A/D} reception,'' in \emph{Proc. IEEE ICASSP}, Seoul, South Korea, Apr. 2024.

\bibitem{Chen2023}
H.~Chen, W.~Wang, W.~Liu, Y.~Tian, and G.~Wang, ``An exact near-field model based localization for bistatic {MIMO} radar with {COLD} arrays,'' \emph{{IEEE} Transactions on Vehicular Technology}, vol.~72, no.~12, pp. 16\,021--16\,030, 2023.

\bibitem{Cheng2023}
C.~Cheng, S.~Liu, H.~Wu, and Y.~Zhang, ``Mixed-field source localization based on robust matrix propagator and reduced-degree polynomial rooting,'' \emph{Signal Processing}, vol. 208, p. 108988, jul 2023.

\bibitem{Xue2023}
D.~Xue, Y.~Guo, L.~Yu, J.~Huo, H.~Chen, and W.~Liu, ``Three-dimensional near-field localization with cross array considering amplitude attenuation,'' \emph{Circuits, Systems, and Signal Processing}, vol.~42, no.~7, pp. 4401--4414, jan 2023.

\bibitem{Gavras_SPAWC_2024}
I.~Gavras and G.~C. Alexandropoulos, ``Simultaneous near-field {THz} communications and sensing with full duplex metasurface transceivers,'' in \emph{Proc. IEEE Int. Workshop Signal Process Adv. Wireless Commun.}, Lucca, Italy, Sep. 2024.

\bibitem{Ebadi2024}
Z.~Ebadi, A.~M. Molaei, M.~A.~B. Abbasi, S.~Cotton, A.~Tukmanov, and O.~Yurduseven, ``Electromagnetic informed data model considerations for near-field {DOA} and range estimates,'' \emph{Scientific Reports}, vol.~14, no.~1, Jul. 2024.

\bibitem{Friedlander2019}
B.~Friedlander, ``Localization of signals in the near-field of an antenna array,'' \emph{{IEEE} Transactions on Signal Processing}, vol.~67, no.~15, pp. 3885--3893, aug 2019.

\bibitem{Ebadi2024a}
Z.~Ebadi, A.~M. Molaei, M.~A.~B. Abbasi, S.~Cotton, A.~Tukmanov, and O.~Yurduseven, ``Near-field localization with an exact propagation model in presence of mutual coupling,'' \emph{arXiv preprint arXiv:2407.19597}, 2024.

\bibitem{Mohsen2023}
N.~Mohsen, A.~Hawbani, X.~Wang, B.~Bairrington, L.~Zhao, and S.~Alsamhi, ``New array designs for doa estimation of non-circular signals with reduced mutual coupling,'' \emph{IEEE Transactions on Vehicular Technology}, vol.~72, no.~7, pp. 8313--8328, 2023.

\bibitem{Lan2023}
L.~Lan, M.~Rosamilia, A.~Aubry, A.~De~Maio, and G.~Liao, ``Adaptive target detection and doa estimation with uniform rectangular arrays in the presence of unknown mutual coupling,'' \emph{IEEE Transactions on Radar Systems}, vol.~1, pp. 325--338, 2023.

\bibitem{Liu2023}
L.~Liu, H.~Zhang, L.~Lan, and J.-Y. Deng, ``Joint range and angle estimation by fda-mimo radar with unknown mutual coupling,'' \emph{IEEE Transactions on Aerospace and Electronic Systems}, vol.~59, no.~4, pp. 3669--3683, Aug. 2023.

\bibitem{Ge2020}
Q.~Ge, Y.~Zhang, and Y.~Wang, ``A low complexity algorithm for direction of arrival estimation with direction-dependent mutual coupling,'' \emph{IEEE Communications Letters}, vol.~24, no.~1, pp. 90--94, Jan. 2020.

\bibitem{Qi2019}
D.~Qi, M.~Tang, S.~Chen, Z.~Liu, and Y.~Zhao, ``Doa estimation and self-calibration under unknown mutual coupling,'' \emph{Sensors}, vol.~19, no.~4, p. 978, Feb. 2019.

\bibitem{Elbir2017}
A.~M. Elbir, ``Direction finding in the presence of direction-dependent mutual coupling,'' \emph{IEEE Antennas and Wireless Propagation Letters}, vol.~16, pp. 1541--1544, 2017.

\bibitem{Zheng2021}
Z.~Zheng and C.~Yang, ``Direction-of-arrival estimation of coherent signals under direction-dependent mutual coupling,'' \emph{IEEE Communications Letters}, vol.~25, no.~1, pp. 147--151, Jan. 2021.

\bibitem{Famoriji2021}
O.~J. Famoriji and T.~Shongwe, ``Source localization of em waves in the near-field of spherical antenna array in the presence of unknown mutual coupling,'' \emph{Wireless Communications and Mobile Computing}, vol. 2021, pp. 1--14, Dec. 2021.

\bibitem{Abedin2012}
M.~J. Abedin and A.~S. Mohan, ``A subspace-based compensation method for the mutual coupling in concentric circular ring arrays for near-field source localisation,'' \emph{International Journal of Antennas and Propagation}, vol. 2012, pp. 1--13, 2012.

\bibitem{Xie2016}
J.~Xie, H.~Tao, X.~Rao, and J.~Su, ``Localization of mixed far-field and near-field sources under unknown mutual coupling,'' \emph{Digital Signal Processing}, vol.~50, pp. 229--239, Mar. 2016.

\bibitem{Wen2023}
K.~Wen, Y.~Tian, and Z.~Dong, ``Mixed source localization considering mutual coupling and unknown nonuniform noise under exact spatial geometry,'' \emph{Signal Processing}, vol. 210, p. 109066, Sep. 2023.

\bibitem{Chen2019}
H.~Chen, W.~Liu, W.-P. Zhu, M.~Swamy, and Q.~Wang, ``Mixed rectilinear sources localization under unknown mutual coupling,'' \emph{Journal of the Franklin Institute}, vol. 356, no.~4, pp. 2372--2394, Mar. 2019.

\bibitem{Famoriji2022}
O.~J. Famoriji and T.~Shongwe, ``Critical review of basic methods on doa estimation of em waves impinging a spherical antenna array,'' \emph{Electronics}, vol.~11, no.~2, p. 208, Jan. 2022.

\bibitem{Lu2023}
Z.~Lu, Y.~Han, S.~Jin, and M.~Matthaiou, ``Near-field localization and channel reconstruction for elaa systems,'' \emph{IEEE Transactions on Wireless Communications}, pp. 1--1, 2023.

\bibitem{NF_beam_tracking}
P.~Gavriilidis and G.~C. Alexandropoulos, ``Near-field beam tracking with extremely massive dynamic metasurface antennas,'' \emph{arXiv preprint arXiv:2406.01488}, 2024.

\bibitem{Boyer2008}
R.~Boyer and G.~Bouleux, ``Oblique projections for direction-of-arrival estimation with prior knowledge,'' \emph{IEEE Transactions on Signal Processing}, vol.~56, no.~4, pp. 1374--1387, Apr. 2008.

\bibitem{CSTStudio}
C.~M. Studio, \emph{{CST} Microwave studio}.\hskip 1em plus 0.5em minus 0.4em\relax CST Studio Suite, 2008.

\bibitem{Liao2012}
B.~Liao, Z.-G. Zhang, and S.-C. Chan, ``{DOA} estimation and tracking of {ULAs} with mutual coupling,'' \emph{IEEE Transactions on Aerospace and Electronic Systems}, vol.~48, no.~1, pp. 891--905, Jan. 2012.

\bibitem{Zhang2018}
X.~Zhang, W.~Chen, W.~Zheng, Z.~Xia, and Y.~Wang, ``Localization of near-field sources: A reduced-dimension {MUSIC} algorithm,'' \emph{{IEEE} Communications Letters}, vol.~22, no.~7, pp. 1422--1425, jul 2018.

\bibitem{Bazzi2016}
A.~Bazzi, D.~T.~M. Slock, and L.~Meilhac, ``Online angle of arrival estimation in the presence of mutual coupling,'' in \emph{2016 IEEE Statistical Signal Processing Workshop (SSP)}.\hskip 1em plus 0.5em minus 0.4em\relax IEEE, Jun. 2016.

\bibitem{Weiss1993}
A.~Weiss and B.~Friedlander, ``Range and bearing estimation using polynomial rooting,'' \emph{IEEE Journal of Oceanic Engineering}, vol.~18, no.~2, pp. 130--137, Apr. 1993.

\bibitem{Grosicki2005}
E.~Grosicki, K.~Abed-Meraim, and Y.~Hua, ``A weighted linear prediction method for near-field source localization,'' \emph{{IEEE} Transactions on Signal Processing}, vol.~53, no.~10, pp. 3651--3660, oct 2005.

\bibitem{Bayesian_CRB_ISAC}
N.~Su, F.~Liu, C.~Masouros, G.~C. Alexandropoulos, Y.~Xiong, and Q.~Zhang, ``Secure {ISAC MIMO} systems: Exploiting interference with {B}ayesian {C}ramér-rao bound optimization,'' \emph{arXiv preprint arXiv:2401.16778}, 2024.

\bibitem{kay}
S.~K. Sengupta and S.~M. Kay, ``\BIBforeignlanguage{eng}{Fundamentals of statistical signal processing: Estimation theory},'' \emph{\BIBforeignlanguage{eng}{Technometrics}}, vol.~37, no.~4, pp. 465--, 1995.

\bibitem{Huang1991}
Y.-D. Huang and M.~Barkat, ``Near-field multiple source localization by passive sensor array,'' \emph{{IEEE} Transactions on Antennas and Propagation}, vol.~39, no.~7, pp. 968--975, jul 1991.

\bibitem{Larmour2024}
C.~Larmour, N.~Buchanan, V.~Fusco, and M.~Ali Babar~Abbasi, ``Sparse array mutual coupling reduction,'' \emph{IEEE Open Journal of Antennas and Propagation}, vol.~5, no.~1, pp. 201--216, Feb. 2024.

\end{thebibliography}
\vspace{12pt}

\end{document}